\documentclass[lettersize,journal]{IEEEtran}
\usepackage{amsmath,amsfonts}
\usepackage{algorithmic}
\usepackage{algorithm}
\usepackage{array}
\usepackage{textcomp}
\usepackage{stfloats}
\usepackage{url}
\usepackage{verbatim}
\usepackage{graphicx}
\usepackage{cite}
\usepackage{comment}
\usepackage{soul}
\usepackage[font=small]{caption}
\usepackage{subcaption}
\usepackage{enumitem}
\usepackage[hidelinks]{hyperref}
\usepackage{tikz}
\usepackage{cuted}
 
\newcommand*\circled[1]{\tikz[baseline=(char.base)]{
            \node[shape=circle,draw,inner sep=2pt] (char) {#1};}}

\begin{document}

\title{Field-Programmable Gate Array Architecture for Deep Learning: Survey \& Future Directions}

\author{Andrew Boutros,~\IEEEmembership{Member,~IEEE,}
        Aman Arora,~\IEEEmembership{Member,~IEEE,}
        Vaughn Betz~\IEEEmembership{Fellow,~IEEE,}
\thanks{
\textbf{A. Boutros} and \textbf{V. Betz} are with the Department of Electrical and Computer Engineering, University of Toronto, Toronto, ON M5S 3G4, Canada (e-mails: andrew.boutros@mail.utoronto.ca; vaughn@eecg.utoronto.ca).
\textbf{A. Arora} is with the School of Computing and Augmented Intelligence, Arizona State University, Tempe, AZ 85281, USA (e-mail: aman.kbm@asu.edu).}}



\begin{figure*}[!t]
\begin{large}
This work has been submitted to the IEEE for possible publication. Copyright may be transferred without notice, after which this version may no longer be accessible.
\end{large}
\end{figure*}
\pagebreak

\markboth{Under review for publication in Proceedings of the IEEE}%
{Boutros \MakeLowercase{\textit{et al.}}: FPGA Architecture for Deep Learning}

\maketitle

\begin{abstract}
Deep learning (DL) is becoming the cornerstone of numerous applications both in large-scale datacenters and at the edge.
Specialized hardware is often necessary to meet the performance requirements of state-of-the-art DL models, but the rapid pace of change in DL models and the wide variety of systems integrating DL make it impossible to create custom computer chips for all but the largest markets.
Field-programmable gate arrays (FPGAs) present a unique blend of reprogrammability and direct hardware execution that make them suitable for accelerating DL inference. 
They offer the ability to customize processing pipelines and memory hierarchies to achieve lower latency and higher energy efficiency compared to general-purpose CPUs and GPUs, at a fraction of the development time and cost of custom chips.
Their diverse and high-speed IOs also enable directly interfacing the FPGA to the network and/or a variety of external sensors, making them suitable for both datacenter and edge use cases.

As DL has become an ever more important workload, FPGA architectures are evolving to enable higher DL performance.
In this article, we survey both academic and industrial FPGA architecture enhancements for DL.
First, we give a brief introduction on the basics of FPGA architecture and how its components lead to strengths and weaknesses for DL applications. 
Next, we discuss different styles of DL inference accelerators on FPGAs that achieve state-of-the-art performance and productive development flows, ranging from model-specific dataflow styles to software-programmable overlay styles.
We survey DL-specific enhancements to traditional FPGA building blocks including the logic blocks, arithmetic circuitry, and on-chip memories, as well as new DL-specialized blocks that integrate into the FPGA fabric to accelerate tensor computations. Finally, we discuss hybrid devices that combine processors and coarse-grained accelerator blocks with FPGA-like interconnect and networks-on-chip, and highlight promising future research directions.
\end{abstract}

\begin{IEEEkeywords}
FPGA, architecture, deep learning, acceleration
\end{IEEEkeywords}

\section{Introduction}
\label{sec:intro}

For many years, computing machines were used to solve problems by efficiently executing sequences of simple and repetitive operations at very high speeds.
A human would think of an algorithmic approach to solve a given problem and then use a programming language to precisely describe its steps.
However, some tasks that are easy for the human brain to perform are very difficult to describe algorithmically.
Detecting a human in an image is one example of such a task.
The high dimensionality of the input and the large number of variations in body pose, size within the image, clothing, image-capture angle, and lighting conditions make it impossible to formulate a set of conditions whose satisfaction implies a human is in the image.
Therefore, solving such tasks classically required a domain expert to hand-craft a set of (lower dimensional) features that can be extracted from the high-dimensional input.
Then, these features are used to train a statistical classifier with many examples and their corresponding output predictions~\cite{dalal2005histograms}.
This approach, typically referred to as classical \emph{machine learning} (ML), requires designing a new feature extractor for each use case and its achieved accuracy is highly dependent on how well these hand-crafted features capture the key relevant data patterns from the original high-dimensional input.

\begin{figure}[!t]
\centering
\includegraphics[width=\linewidth]{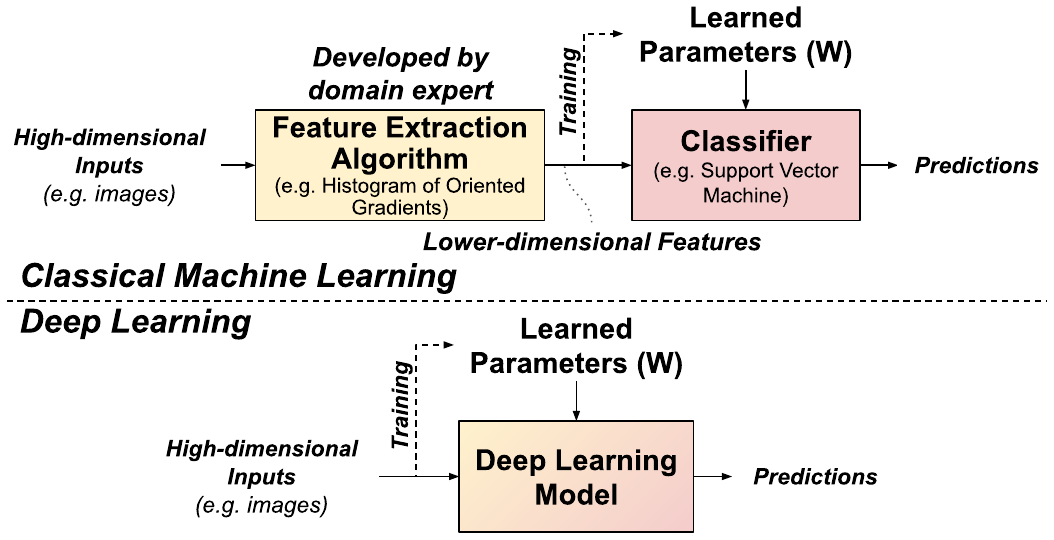}
\caption{Comparison between the classic ML approach using hand-crafted features to train a statistical classifier (top), and DL models trained directly on input data to perform both feature extraction and classification (bottom).}
\label{fig:ml-dl}
\end{figure}

\IEEEpubidadjcol
With the abundance of training data and the continuous improvement of compute capabilities in recent years, it became feasible to train large artificial neural networks using theory and algorithms that have been formulated back in the 1980s~\cite{rumelhart1988learning}.
These neural networks are typically composed of a deep cascade of layers, and therefore are referred to as \emph{deep learning} (DL) models. 
Each layer contains a number of \emph{neurons} performing a weighted sum of their inputs. 
As illustrated in Fig.~\ref{fig:ml-dl}, the key distinction from classical ML methods is that DL models can learn to extract features and classify them directly from the input training data instead of relying on features hand-crafted by a domain expert, resulting in better accuracy and wider applicability to different domains.
Since deep neural networks (DNNs) first demonstrated their superior quality of results in 2012 on visual recognition and image classification tasks~\cite{krizhevsky2012imagenet}, there has been an avalanche of innovations in building better DL models that can achieve higher accuracy in many different domains such as natural language processing (NLP)~\cite{vaswani2017attention}, recommendation systems~\cite{naumov2019deep}, and content generation~\cite{ramesh2021zero}.
Nowadays, DL models enable a myriad of day-to-day end user applications~\cite{haldar2019applying, capes2017siri, steck2021deep}, facilitate software development~\cite{chen2021evaluating}, boost computer chip design productivity~\cite{liu2023chipnemo}, and push the boundaries of human knowledge by discovering more efficient computational algorithms~\cite{fawzi2022discovering} and solving long-standing scientific problems~\cite{jumper2021highly}.

However, this comes at the cost of a significantly higher computational complexity and memory footprint compared to classical ML methods~\cite{suleiman2017towards}.
Google has estimated that if their users perform voice search for only 3 minutes per day using DL-based speech recognition on general-purpose CPUs, it would require doubling their datacenters' compute capacity~\cite{jouppi2017datacenter}.
A recent report~\cite{patel2023inference} estimates that ChatGPT, OpenAI's conversational DL model, costs around \$0.7M in compute hardware costs per day to serve a small fraction of queries compared to those processed by the Google search engine.
These DL compute demands are rapidly growing, challenging the capabilities of conventional general-purpose compute platforms.
Therefore, application-specific (ASIC) accelerators are deployed in both datacenters and edge devices to increase the efficiency of DL computation~\cite{jouppi2017datacenter}.
In addition, with DL being such a pervasive workload, it is also driving architectural innovations in almost all forms of general-purpose compute platforms to improve their DL compute efficiency.
For example, Intel's fourth generation Xeon (Sapphire Rapids) central processing units (CPUs) support more efficient DL-targeted tensor instructions~\cite{khaldi2021extending} and next-generation AMD Ryzen 7000 processors are integrating acceleration engines for artificial intelligence (AI) workloads~\cite{amdkeynote}.
Modern graphics processing units (GPUs) also include specialized tensor cores to improve the efficiency of the matrix multiplications extensively used in DL workloads.
While tensor operations are key in DL, they are not the entire compute pipeline and bottlenecks can still occur elsewhere. 
As an example, Nvidia recently integrated dedicated engines for DL preprocessing operations such as JPEG image decoders in their A100 GPUs to address one such bottleneck~\cite{weissenberger2021accelerating}.

The architecture of field-programmable gate arrays (FPGAs) is similarly driven by their key use cases.
Therefore, we are also starting to witness many FPGA architecture changes targeted at making them more efficient for DL.
In this article, we survey proposals and innovations from both academia and industry to optimize FPGA architecture specifically for DL.
We first present an introductory tutorial on FPGA architecture through a DL lens, highlighting the key strengths, weaknesses, and opportunities of FPGAs in the area of DL acceleration.
Then, we highlight the key design styles of DL accelerators on FPGAs with selected examples from the broad literature available on this topic.
However, this article is not intended to be a comprehensive survey of FPGA-based DL accelerator implementations and toolflows.
We refer interested readers to~\cite{guo2019dl, abdelouahab2018accelerating, venieris2018toolflows} which cover this area in more detail.
Next, we explain the general methodology used to model and evaluate new FPGA architectures.
We discuss DL-driven architecture enhancements in modern FPGAs, starting from conventional FPGA blocks (e.g. logic elements and embedded hard blocks) and moving on to new DL-specific blocks (e.g. tensor blocks) as well as on-die coarse-grained accelerators (e.g. AI engines) and in-package DL chiplets.
Finally, we present our perspective on the future of reconfigurable devices in the DL domain and identify interesting research directions in this area.
\section{FPGA for DL Acceleration}
\label{sec:fpga_for_dl}

\subsection{Key DL Acceleration Requirements}
\label{sec:dl_reqs}

\begin{figure}[!t]
\centering
\includegraphics[width=\linewidth]{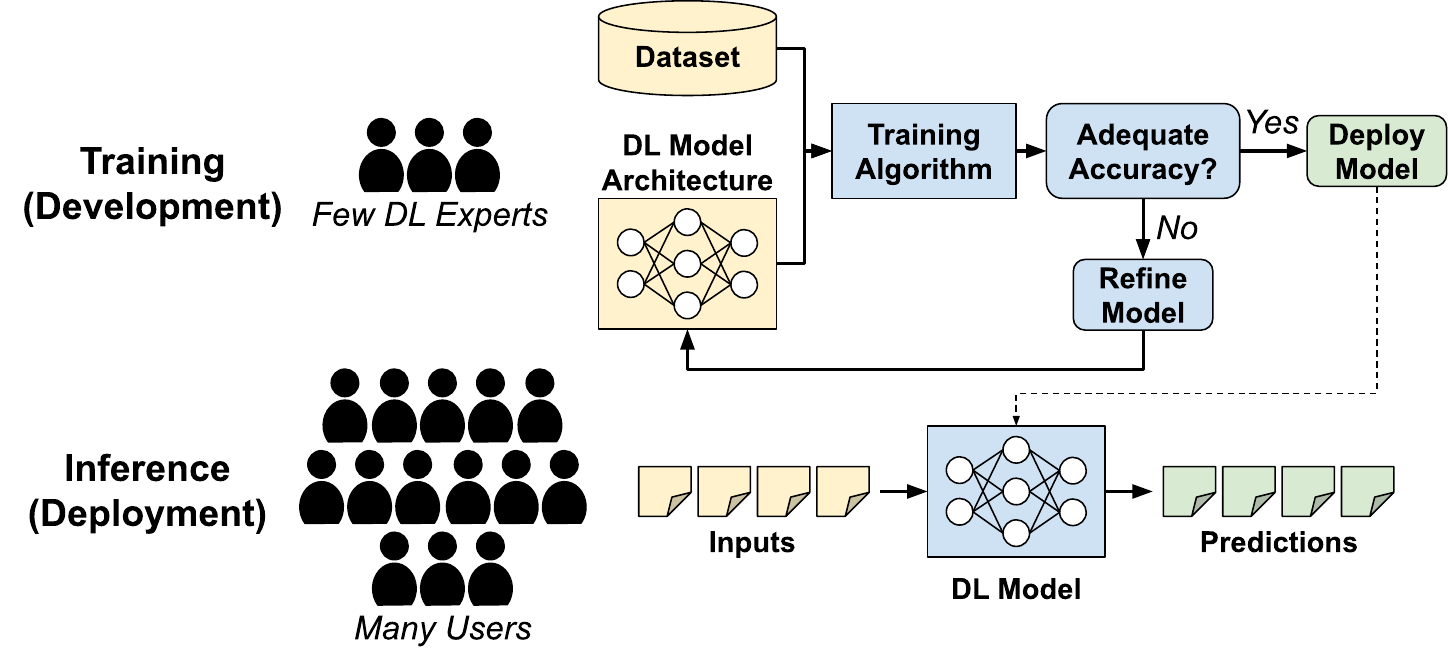}
\caption{The training and inference phases of a DL model. Training is performed by a few DL experts on large-scale compute clusters and requires many design iterations. When model accuracy is satisfactory, it is deployed for inference with varying performance requirements depending on the application (latency-tolerant vs. real-time).}
\label{fig:training_inference}
\end{figure}

As illustrated in Fig.~\ref{fig:training_inference}, a DL model goes through two main phases during its life-cycle: training and inference.
During training, a small group of DL scientists architect the model and train it using huge datasets (e.g. $>$570~GB for the GPT-3 language model~\cite{brown2020language}) to achieve the desired quality of results.
This process is very compute/memory intensive due to the large amount of data used for training and usually takes tens to hundreds of design iterations to optimize the model.
Therefore, it is typically performed on large clusters of compute machines and accelerators in a datacenter.
The final product of the training process is a model architecture and values of its trainable parameters or \emph{weights}, which are then deployed in a production-scale application to perform inference on new data samples that were not part of the training dataset.
Depending on the deployment environment (cloud/datacenter vs. edge/embedded) and the nature of the application (latency-tolerant vs. real-time), DL inference can have different compute requirements and constraints. 
Consequently, the hardware used to accelerate DL training and inference has to be optimized for different metrics and use cases, creating potential markets for different acceleration platforms (e.g. GPUs, FPGAs, and ASICs) based on their characteristic strengths and weaknesses.

\subsubsection{Performance}
Performance of DL accelerators is measured using two metrics: throughput and latency.
Throughput is the number of input examples a specific accelerator can process per unit time on a given DL workload. 
To facilitate accelerator efficiency comparisons across models of different computational complexity, throughput is typically reported in giga or tera operations per second (GOPS or TOPS) where the operations are typically computed as the number of multiplies plus the number of accumulates as multiply-accumulate (MAC) is the dominant operation in DL workloads. 
Each accelerator has a \emph{peak throughput} which is workload-independent, and is determined by the number of MACs it can perform per cycle and its maximum operating frequency.
However, in practice it is not possible to achieve 100\% utilization of these MAC units\footnote{Depending on the model architecture and the hardware organization, the MAC units can be idle during some cycles. For example, computation can stall when model weights are being loaded from external memory or transported from on-chip buffers to compute units, and some compute units can be idle when the input size does not exactly match the hardware compute parallelism (e.g. performing a dot product between two 6-element vectors on an 8-lane dot product engine).}, and thus the \emph{effective throughput} of an accelerator is the more realistic metric and is typically evaluated for each DL workload~\cite{boutros2020beyond}.
An efficient accelerator architecture aims to maximize its \emph{compute utilization} (i.e.~minimize the gap between peak and effective throughput). 
To improve their effective throughput, many accelerators \emph{batch} a group of inputs to be processed at the same time.
This enables reusing the same set of model weights across the many inputs in a batch to hide the memory latency of loading the next set of weights and reduces the number of cycles in which the MAC units remain idle.
On the other hand, latency is the amount of time it takes the accelerator to process a single input, which is a key metric for real-time applications.
Although batching can help improve the effective throughput of an accelerator, it typically increases latency since more time is needed to form a batch, process the entire batch of inputs, and output all their results at the same time.
As an example, for the ResNet-50 image classification model, increasing the batch size from 1 to 8 inputs improves throughput by 3$\times$ at the cost of 2.2$\times$ higher latency on an Nvidia V100 GPU~\cite{hall2020tensorflow}.

For the training phase, latency is not a concern, and therefore DL training accelerators are throughput-optimized to maximize the number of training samples that can be processed per second.
For inference, the optimization target depends on the use case. 
Applications such as DL-based image search engines or video copyright checks focus mainly on maximizing  the number of user queries that can be served per second with a loose latency requirement, and therefore are throughput-oriented.
In other applications such as pedestrian or obstacle detection in an autonomous vehicle, the latency of acquiring inputs from several cameras or sensor readings, detecting pedestrians/obstacles using one or multiple cascaded DL models, and then taking an action (e.g. adjusting direction or applying the brakes) is crucial for safety reasons.

\subsubsection{Cost and Energy Efficiency}

For both the training and inference phases, energy and cost efficiency are major design optimization targets for all DL accelerators.
It is estimated that 35\% of the total cost of ownership of a datacenter is spent on power~\cite{ganesh2013integrated}.
Therefore, with DL becoming a prominent datacenter workload, more energy efficient DL compute hardware directly translates to significant cost savings for service providers.
For example, Google reported that using their ASIC tensor processing unit (TPUs) reduced the cost of training a ResNet-50 model for image classification by 38\%~\cite{google2018cost}.
For modern NLP models such as BERT~\cite{devlin2018bert}, this cost can reach millions of dollars for each full training~\cite{sharir2020cost}.
Additionally, reducing the power consumption of datacenter compute has significant environmental impact, as datacenters consume very large amounts of electricity and by some estimates will account for $\sim$8\% of the world's electricity demand by 2030~\cite{jones2018stop}.
At the other end of the deployment spectrum, DL inference on battery-operated edge devices usually has a very tight power budget and therefore requires energy-efficient compute hardware.
For example, Tesla's full self-driving DL inference chip was custom designed to meet an aggressive power budget of less than 40W~\cite{talpes2020compute}.
Although such custom ASICs can offer superior energy-efficiency, they lack the flexibility to adapt to different systems and algorithms.
In addition, their significant non-recurring engineering (NRE) cost and longer time-to-solution for design, fabrication and testing can be prohibitive for small and medium-scale markets.

\subsubsection{Adaptability}
Besides achieving high performance and energy-efficiency, DL accelerators must also be flexible to adapt to frequent algorithmic changes.
New state-of-the-art DL models are introduced at a much faster rate than the typical design and deployment cycle of computer hardware.
Adaptability can be achieved by making some or all of the system software-programmable, but software programmability adds energy and performance overheads (e.g.~instruction fetching and decoding) compared to fixed-function dedicated hardware.
In addition, the DL accelerator, especially in edge deployments, is typically part of a bigger system where it needs to interface with a wide variety of sensors (e.g.~cameras, LiDARs) and actuators (e.g.~to control brakes) which can have different communication protocols as well as different data pre- and post-processing needs.
Therefore, adaptability is not only a requirement for the DL compute hardware, but also for interfacing with the rest of the system to ensure usability across a wide range of deployment use cases. 
Enabling such flexible interfacing often requires electrical adaptability due to the different voltage and timing requirements of different interface protocols, which can not be achieved by software programmability alone.

\subsection{FPGA Strengths for DL Acceleration}
\label{sec:fpga_strengths}

\begin{figure}[!t]
\centering
\includegraphics[width=\linewidth]{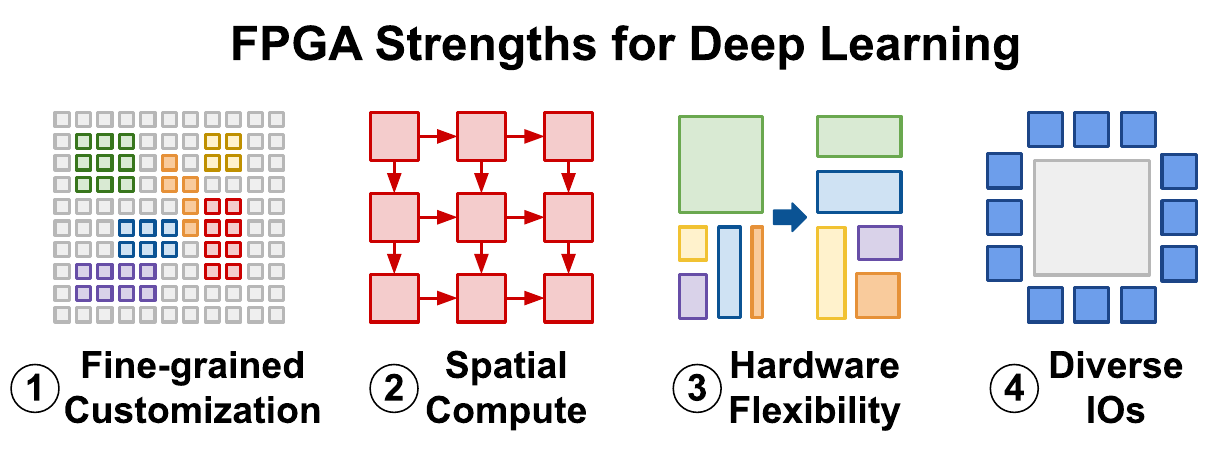}
\caption{Unique features of FPGAs that make them an efficient acceleration platform for DL.}
\label{fig:fpga_strengths}
\end{figure}

Based on the DL acceleration requirements discussed in the previous subsection, we can identify key FPGA strengths (summarized in Fig.~\ref{fig:fpga_strengths}) that make them a desirable and efficient acceleration platform for specific DL use cases.

Firstly, \textbf{FPGAs offer fine-grained hardware programmability} which allows building customized compute datapaths and on-chip memory sub-systems that match exactly the application needs.
Therefore, FPGAs have an advantage compared to general-purpose processors (e.g.~CPUs and GPUs) in use cases where customization is desired.
For example, DL inference quality of results is tolerant of low-precision computation and since the energy and area of compute units drop rapidly with precision, this can be exploited for a more efficient acceleration solution~\cite{darvish2020pushing}. 
Unlike CPUs and GPUs that support only specific arithmetic precisions (e.g.~\texttt{int4, int8, fp16, fp32}), an FPGA can implement custom compute units for any precision, including binary/ternary, narrow floating-point formats (e.g.~\texttt{fp8}~\cite{micikevicius2022fp8}), or floating-point numbers with shared exponents (\texttt{bfp})~\cite{darvish2020pushing}.
Since the most efficient precision varies across DL models and even across layers within a single model, this flexibility is very useful.
However, the fine-grained bit-level hardware programmability comes with speed and area overheads as functions are implemented with programmable blocks and programmable routing which are slower and bigger than standard cell logic gates and direct wires.
Therefore, the customization gains should outweigh these programmability overheads for an FPGA acceleration solution to be competitive.

Secondly, \textbf{FPGAs are spatial computing devices}.
This means that data does not have to move through a memory hierarchy of caches and register files for the computation to be performed, and compute cores do not have to communicate through memory.
Instead, in an FPGA, data can flow directly from distributed on-chip buffers and through \emph{chained} compute units using the flexible programmable routing, without the need for an instruction sequence to orchestrate data movements and computations. 
This can reduce the overall compute latency as fewer cycles are spent on data movement across different levels of the memory hierarchy.
It can also result in significant energy savings; for example, {$\sim$}99\% of the energy consumed by an integer add operation in a 45nm CPU is spent on cache/register file accesses and control logic~\cite{horowitz20141}, a large portion of which can be saved when performing computations spatially on an FPGA.

Thirdly, \textbf{FPGAs are flexible}.
Reconfiguring the FPGA with a new bitstream changes its hardware functionality.
This offers a clear advantage over ASIC accelerators since the hardware can flexibly adapt to rapid changes in DL algorithms, model architectures and application-specific pre- or post-processing.
New operations can be implemented in hardware, integrated into an FPGA-based accelerator architecture, and deployed in production in a matter of weeks~\cite{chungretrospective}.
On the other hand, an ASIC accelerator would need to implement this new operation on a software-programmable core or its host CPU resulting in degraded performance until the next generation chip is designed, fabricated and deployed, which can take years.

Fourthly, \textbf{FPGAs provide a myriad of programmable input/output (IO) interfaces}.
These IOs can flexibly implement a wide variety of protocols with different electrical characteristics and timing specifications. 
Modern FPGAs also implement hardened controllers for various widely-used standards for datacenter deployments such as ethernet, peripheral component interconnect express (PCIe), and double data rate (DDR) and high bandwidth (HBM) external memories.
This allows efficient communication between the FPGA as a server accelerator card and the host CPU, and also enables directly connecting multiple FPGAs over the network to create many-device accelerators such as Microsoft's Brainwave datacenter-scale DL accelerator~\cite{fowers2018configurable}.
Additionally, the FPGA programmable logic can also implement other custom standards for interfacing with different sensors/actuators in embedded systems at the edge~\cite{urbina2019smart}.

These unique FPGA characteristics lead to certain DL use cases where FPGAs have advantages compared to other acceleration solutions such as general-purpose CPUs/GPUs and ASIC accelerators. 
These are use cases that:
\begin{itemize}[noitemsep,topsep=0pt,leftmargin=2\labelsep]
    \item can perform computations using low precision or non-standard number formats which require customized datapaths~\cite{mishra2017wrpn, darvish2020pushing}. These precisions are more common and generally easier to use in inference, while training is typically performed in higher precision floating-point formats (e.g.~\texttt{fp32, fp16}) that are natively supported in general-purpose CPUs and GPUs.
    \item have tight latency constraints that prohibit batching a large number of inputs for processing. If an application is more throughput-oriented with loose latency constraints, multiple inputs can be batched and processed simultaneously. This provides more opportunities for reuse of on-chip values, and while this can benefit all computational devices it is usually particularly helpful  in keeping a GPU's massively parallel functional units busy. This is another reason why (latency-constrained) inference is a better match for FPGAs than (throughput-oriented) training.
    \item can fit all model weights in the FPGA's on-chip memory (i.e.~\emph{persistent weights}). The spatial nature of FPGAs enables near-memory compute with low-latency memory accesses and application-tailored memory organization. For bigger models, the diverse FPGA IOs allow directly connecting multiple FPGAs over the network to create bigger multi-FPGA fabrics with more on-chip memory~\cite{fowers2018configurable}. 
    \item implement a DL component in a bigger system, in which the FPGA's flexibility and rich IOs can play a crucial role, such as in autonomous driving systems. A variety of sensor/camera inputs might require classical signal/image preprocessing before being used as inputs to a DL model, and then the output of the DL model is used to control different actuators. In such cases, FPGAs offer a platform for accelerating the full system with custom interfaces and application-dependent pre- and/or post-processing~\cite{cheng2019dlbooster}.
    \item require periodic changes to the DL model architecture with new operations and irregular compute graphs~\cite{pham2018efficient}. In contrast to an ASIC, these new changes can be implemented in hardware by simply programming the FPGA with a new bitstream. However, if these changes are very frequent (e.g.~daily), compiling a new bitstream every time the model changes can be challenging due to the high runtime of FPGA computer-aided design (CAD) tools. In such cases, a software-programmable solution could be more desirable.
\end{itemize}

\noindent
Thus, for the rest of this article, we mainly focus on DL inference acceleration which better matches the unique FPGA characteristics and strengths.

\subsection{FPGA-Based DL Inference Acceleration Styles}
This subsection presents commonly-used styles for accelerating DL inference on FPGAs.
It is not intended to be a comprehensive survey of DL inference accelerators implemented on FPGAs, as our focus is primarily on enhancements to the underlying FPGA chip architecture.

In 2012, AlexNet was the first convolutional neural network (CNN) to demonstrate the superior quality of results of DL in image classification tasks compared to prior machine-learning-based approaches~\cite{krizhevsky2012imagenet}.
Its significantly higher computational complexity sparked interest in accelerating DL inference using specialized hardware on FPGAs as co-processors.
A host or an embedded CPU would offload the computation of the whole CNN (or specific compute-intensive layers) to an FPGA accelerator, and at the end perform a final softmax operation to calculate prediction probabilities from the final output of the accelerator, if needed.
In this case, the FPGA accelerator is usually \textbf{hand-crafted and optimized for a specific DL model} or a group of similar models~\cite{suda2016throughput, zhang2015optimizing, qiu2016going, li2016high}.
This approach achieved significant performance and energy efficiency gains compared to software-based solutions on contemporaneous multi-core CPUs and GPUs. 
Although the main focus was on CNN acceleration and computer vision applications, several works also investigated FPGA acceleration of other types of DL models such as recurrent neural networks (RNNs) for sequence inputs and natural language processing~\cite{nurvitadhi2016accelerating, chang2015recurrent, li2015fpga}.

However, with the continuous and rapid advances in state-of-the-art DL models, it quickly became evident that building a custom FPGA accelerator for each model is extremely laborious and cannot keep pace with DL model evolution.
Therefore, building \textbf{custom hardware generators} to automate this process became a major research focus.
These hardware generators are domain-specific compilers that take as inputs the specifications of a target FPGA and the dataflow graph of a DL model in the same formats used by common DL frameworks such as TensorFlow~\cite{abadi2016tensorflow} or PyTorch~\cite{paszke2019pytorch}.
They optimize the input dataflow graph by reordering, simplifying and/or fusing model layers/operations, and then use a library of parameterized implementations of hardware modules to generate an optimized model-specific FPGA implementation given the target FPGA resource constraints.
Although these DL hardware generation toolflows all share the same fundamental concepts, their generated accelerator architectures can be very different.
Some toolflows generate streaming layer-pipelined architectures in which each layer has a dedicated compute engine and all layers coexist on the FPGA in a pipelined fashion (i.e.~\emph{spatial} execution).
Examples of such toolflows are HPIPE~\cite{hall2020tensorflow}, fpgaConvNet~\cite{venieris2018fpgaconvnet}, DNNBuilder~\cite{zhang2018dnnbuilder}, and FINN~\cite{umuroglu2017finn}.
Other toolflows generate architectures that have a number of more flexible processing elements (PEs) on which layers of a given model are mapped and executed sequentially (i.e.~\emph{temporal} execution) as orchestrated by control finite-state machines and microcodes~\cite{ma2016scalable, sharma2016high, guan2017fp, abi2023gnnbuilder}.
Many of these toolflows automatically apply different FPGA-friendly DL model optimizations to further enhance performance such as quantizing to lower numerical precisions and exploiting sparsity to skip ineffectual computations with zero weights. Some work has gone even further, and directly uses the look-up tables (LUTs) in FPGA fabrics as the trainable building blocks of a neural network, instead of using only MAC  operations~\cite{wang2020lutnet, andronic2023polylut}.

The generation of custom DL hardware exploits the unique reconfigurable nature of FPGAs by optimizing the accelerator architecture to exactly match the needs of a specific model or class of models, minimizing the additional overheads of generality.  
However, this comes at the cost of (1)~long FPGA compile time (on the order of hours for synthesis, place and route) to generate a different FPGA bitstream for each new model or slight change in an existing model and (2)~the~need to reprogram the FPGA (which can take tens to hundreds of milliseconds) when switching between pre-compiled bitstreams implementing different models.
These drawbacks can be prohibitive for use cases in which very frequent (e.g.~daily) model updates are deployed in production or when real-time or input-dependent switching between several models is needed.
Another approach for designing DL accelerators on FPGAs is to build \textbf{software-programmable domain-specific overlays}.
Similarly to CPUs and GPUs, an FPGA overlay defines an instruction set architecture (ISA) that decouples the hardware and software stacks. 
The ISA abstracts away all the micro-architecture and hardware implementation details from application developers who write their algorithms in a high-level programming language and then compile them into sequences of instructions that can run on any processor that supports the same ISA. 
For a generic processor architecture and ISA (e.g. RISC-V), a \emph{hard} ASIC implementation will always be more efficient than an FPGA overlay due to the overhead of reconfigurability~\cite{wong2011comparing,boutros2018you}. 
However, a soft processor can enhance efficiency by exploiting the FPGA's flexibility to implement a customized datapath and memory hierarchy, as well as a domain-specific ISA.

\begin{figure}[!t]
\centering
\includegraphics[width=0.87\linewidth]{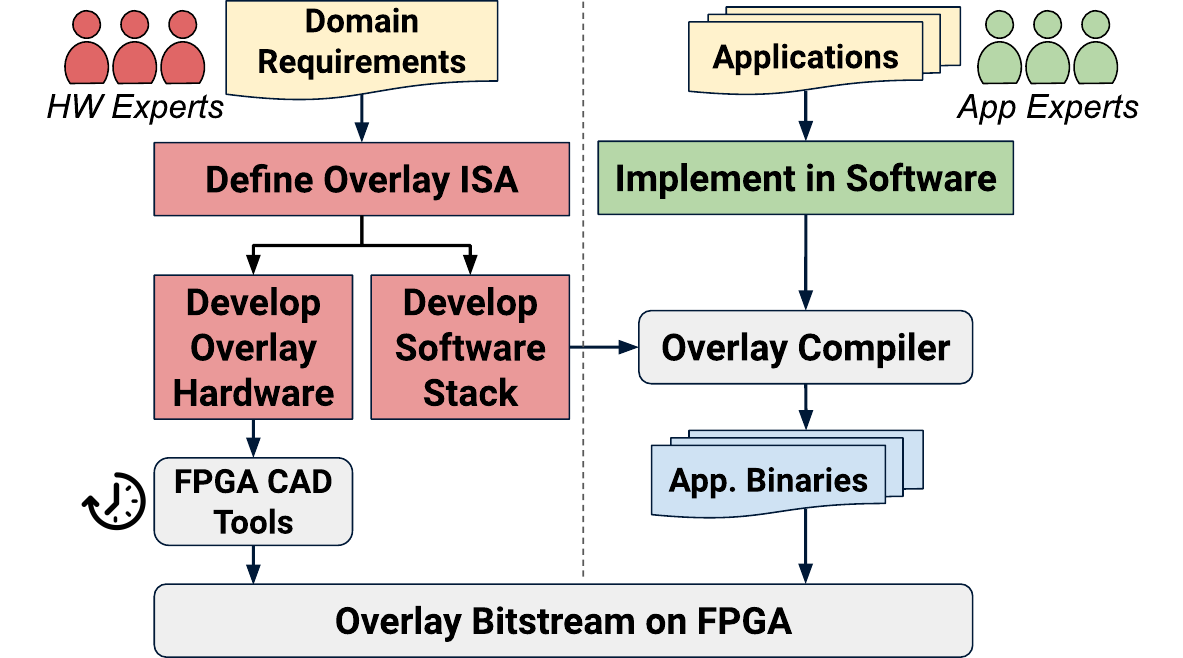}
\caption{The overlay design approach enables application experts to use FPGAs for accelerating their DL workloads without any hardware design expertise or suffering from long runtime of FPGA CAD tools.}
\label{fig:overlay_flow}
\end{figure}

As illustrated in Fig.~\ref{fig:overlay_flow}, to build an FPGA DL overlay, architects would first design the overlay ISA and processor architecture.
The microarchitecture of the overlay is then heavily-optimized to generate a single high-quality FPGA implementation that is deployed on an FPGA and programmed through software to execute different DL models.
To program the overlay, users are provided with a compiler that translates a high-level description of a DL model (e.g.~TensorFlow or PyTorch) to a sequence of instructions to be executed on the FPGA overlay.
In this approach, users do not need to have any hardware design expertise, significantly reducing the barrier of entry for DL application developers to use FPGAs. 
In addition, their iteration time to compile a new DL model is much faster, as they are performing a fast software compile (seconds) to create a new sequence of overlay instructions instead of a long FPGA hardware compile (hours) to create a new bitstream.
There are many DL overlay examples from both industry and academia optimized for different types of models~\cite{aydonat2017opencl, abdelfattah2018dla, xilinxdpu, boutros2020beyond, yu2019opu, bai2023transformer, hur2023fast}, including Microsoft's datacenter-scale DL inference accelerator, Brainwave~\cite{fowers2018configurable}.

\subsection{Examples of DL Acceleration on FPGAs}

When using FPGAs for DL inference acceleration, regardless of the accelerator design style, there are two main concerns.
The first is ease of use; FPGAs are generally harder to design for, use and debug compared to other compute platforms such as GPUs and CPUs.
Even with advances in high-level synthesis (HLS) technology, using FPGAs still requires extensive hardware design expertise, making them harder to use by DL application developers.
The second concern is whether FPGAs can deliver state-of-the-art DL inference performance despite their inherent overhead of reconfigurability.
As discussed in the previous subsection, both the custom hardware generation and overlay design approaches address the first concern.
Using these approaches, DL application developers can go from a high-level DL model description to an FPGA deployment with little or no hardware design expertise. 
In this subsection, we cover two examples from these two design approaches to showcase that FPGAs can deliver best-in-class DL inference performance.
We also show that even higher performance can be realized by optimizing the underlying FPGA architecture specifically for DL.

\subsubsection{Custom Hardware Generation Example (HPIPE)}

\begin{figure}[!t]
\centering
\includegraphics[width=\linewidth]{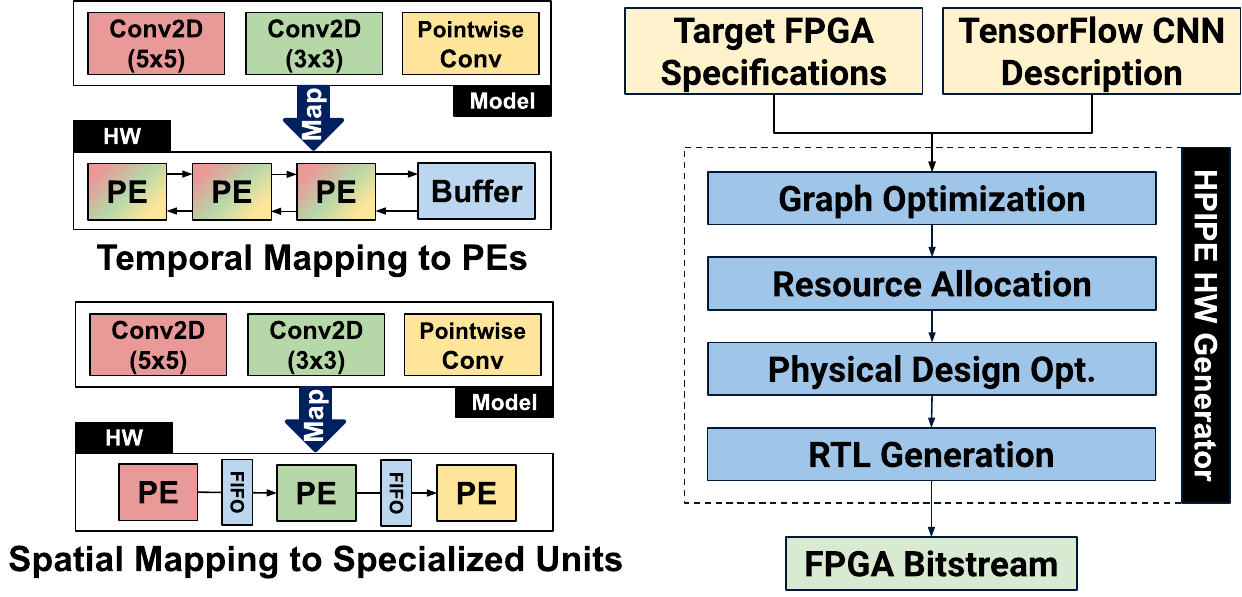}
\caption{Temporal mapping of DL models to an array of PEs (top left) vs. building per-layer customized compute units as in HPIPE (bottom left) and an overview of the HPIPE hardware generation flow (right).}
\label{fig:hpipe_flow}
\end{figure}

HPIPE~\cite{hall2020tensorflow} is a domain-specific compiler that generates layer-pipelined dataflow FPGA accelerators for persistent CNNs, where all the weights can be stored in the on-chip SRAMs.
It builds a unique processing module for each layer in a CNN and chains them using latency-insensitive FIFOs. 
It also exploits weight sparsity by skipping ineffectual zero-weight computations, which can significantly reduce on-chip memory requirements and improve performance by executing fewer operations.
When compared to the one-size-fits-all approach in which the same array of PEs is used for all layers (see left side of Fig.~\ref{fig:hpipe_flow}), the per-layer customized modules in HPIPE result in better utilization of the compute resources and exploit the additional dimension of pipeline parallelism in which all the CNN layers are executing simultaneously on different parts of an image or on different images.   
As illustrated on the right side of Fig.~\ref{fig:hpipe_flow}, the HPIPE compiler takes as inputs a TensorFlow description of the CNN and specifications of the target FPGA.
Then, it performs a number optimizations on the CNN dataflow graph (e.g. fuse layers for a more efficient implementation).
After that, it allocates hardware resources to each CNN layer to balance the throughput of all the pipelined layers and maximize overall performance.
The HPIPE compiler also performs several physical design optimizations that consider the spatial layout of the layer modules and implements optimized interconnect structures for high-fanout and long-span connections, resulting in high operating frequencies.
Finally, it generates the accelerator RTL files and memory initialization files to store the CNN weights in the on-chip memories; the resulting RTL is compiled to a bitstream using the conventional FPGA CAD tools.

Using an Intel Stratix 10 GX2800, the largest monolithic (single die) 14nm Stratix 10 FPGA, HPIPE outperforms all other FPGA-based CNN accelerators on same-generation FPGAs.
It can also achieve 4$\times$ higher ResNet-50 throughput at the same latency ($<1$ms) compared to batch-1 inference on the Nvidia V100 GPU, which is on a comparable process techology (12nm). 
Increasing the batch size improves GPU utilization but worsens latency; HPIPE still achieves 1.4$\times$ higher throughput but at 2.2$\times$ lower latency compared to the V100 GPU running at a batch size of 8.
This highlights the utility of FPGAs for low-latency inference; in this case the FPGA's flexibility enables extreme per-model customization, yielding efficiency gains that outweigh its inherent reconfigurability overheads.
The automatic generation of hardware from a high-level model description eliminates the need for FPGA design expertise, but it still requires a time-consuming compilation of a new FPGA bitstream for each different model to be deployed.

\subsubsection{Overlay Example (NPU)}

\begin{figure*}[!t]
\centering
\includegraphics[width=0.9\linewidth]{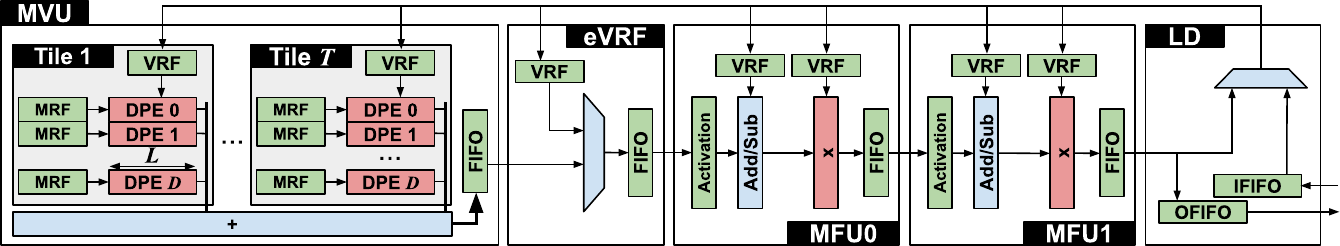}
\caption{The NPU overlay architecture. It is a VLIW processor consisting of five chained coarse-grained units: a matrix-vector multiplication unit (MVU), an external vector register file (eVRF), two multi-function units (MFUs) for vector elementwise operations, and a loader unit.}
\label{fig:npu_arch}
\end{figure*}

The neural processing unit (NPU)~\cite{nurvitadhi2019compete} is a very-long-instruction-word (VLIW) processor architecture targeting low-latency batch-1 inference of DL models with no data reuse (i.e. memory-bound) such as  different types of RNNs and multi-layer perceptrons (MLPs).
The NPU overlay design relies on two key principles.
First, it exploits the massive parallelism of DL models to amortize the energy and area cost of software programmability. 
A single coarse-grained VLIW instruction can trigger the execution of thousands of operations, much like an extreme example of a complex instruction set computer (CISC) architecture.
Second, it customizes the processor's memory subsystem to utilize the tremendous bandwidth between the distributed FPGA on-chip memories and processing elements, performing near-memory compute. 
The memory subsystem is explicitly managed (no caches), uses several different register files with specific purposes and wide data words rather than one general purpose group, and directly chains many operations between functional units with no register file access.
Fig.~\ref{fig:npu_arch} shows the architecture of the NPU overlay, which consists of five coarse-grained chained units such that the output of one unit feeds the next. 
The pipeline starts with a matrix-vector multiplication unit (MVU). 
The MVU consists of $T$ compute tiles, each of which has $D$ dot-product engines (DPEs) of size $L$ multiplication lanes. 
Vector operands are broadcast from a vector register file (VRF) to all DPEs in a single tile, while persistent model weights come from the matrix register files (MRFs). 
The MVU is followed by an external VRF (eVRF) that enables skipping the MVU if an instruction does not start with a matrix-vector multiplication. 
The rest of the pipeline consists of two multi-function units (MFUs) that implement vector elementwise operations (e.g. activation functions, addition, multiplication), and a loader unit which can write back results to any of the processor architecture states (i.e. VRFs) or communicate with external components (e.g. a network interface) through input/output FIFOs. 
DL application developers describe their models using a subset of the Tensorflow Keras API~\cite{keras} which is then compiled into a sequence of NPU VLIW instructions to be executed on the FPGA overlay.

The NPU implemented on an Intel Stratix 10 GX2800 FPGA achieves 2.7$\times$ lower batch-1 latency than the equivalent Nvidia V100 GPU for various RNN workloads from the DeepBench suite~\cite{deepbench} when using the same \texttt{fp32} numerical precision as the GPU.
When using the more FPGA-friendly 8-bit integer precision, this performance gap grows to 8.6$\times$.
This shows that a domain-specific FPGA overlay with a custom architecture and ISA can deliver significantly higher performance compared to generic processor pipelines such as those of GPUs and CPUs, while providing similar software-level programmability.

\subsubsection{Effect of FPGA Architecture Enhancements for DL}
\label{sec:tb_effect}

Both HPIPE and the NPU overlay were designed to best match the underlying FPGA architecture.
For example, they both organize their fundamental MAC compute units to efficiently utilize the embedded digital signal processing blocks (DSPs) in the target FPGA.
HPIPE used the dedicated (non-programmable) interconnects between DSP blocks to build efficient pipelined dot products with minimal utilization of the FPGA's programmable logic and routing.
On the other hand, the NPU used a small amount of soft logic for post-multiplier correction to enable the dense packing of four \texttt{int8} multipliers to the two \texttt{int18} multipliers available in a single DSP block on an Intel Stratix 10 FPGA~\cite{langhammer2019extracting}.
These optimizations are used to enhance DL performance assuming that the FPGA architecture itself is a constant.
However, FPGA architecture has been continuously evolving to better suit key FPGA use cases and market segments throughout the past three decades.
Thus, with DL becoming such a prominent workload, many DL-targeted FPGA architecture enhancements have been proposed in the past few years.

One example of such FPGA architecture enhancements (which we cover in more detail later in this article) is the replacement of conventional DSP blocks with DL-optimized tensor blocks in the Intel Stratix 10 NX FPGA~\cite{langhammer2021stratix}.
These tensor blocks replace the legacy DSP block modes of operation and precisions with DL-targeted ones that can implement significantly more \texttt{int8} and \texttt{int4} multiplications per block. By restricting the data input and output patterns that can achieve peak throughput (and thereby avoiding adding expensive additional programmable interconnect), these tensor blocks achieve area similar to that of a conventional DSP block.
Both HPIPE and the NPU were upgraded to use these new DL-optimized FPGAs with tensor blocks resulting in a significant performance boost.
For HPIPE, the tensor blocks improved inference throughput by 4.8$\times$ from 6,000 to 29,400 batch-1 inferences per second for the MobileNet-V2 CNN compared to the conventional FPGA with DSP blocks~\cite{stan2022hpipe}.
Compared to the Nvidia V100 GPU (which is more than 1.5$\times$ bigger in die size than the Stratix 10 NX FPGA), the tensor-block-enhanced HPIPE achieves 17$\times$ higher throughput and 3$\times$ lower latency at batch-1 or 1.3$\times$ higher throughput and 29$\times$ lower latency at batch-128.
On the other hand, the NPU performance is improved by 3.5$\times$ when using the tensor blocks compared to conventional DSPs, resulting in 11$\times$ higher performance than the V100 GPU~\cite{boutros2020beyond}.

These two examples highlight the significant impact DL-specific architecture enhancements can have on FPGA inference performance.
In this article, we cover many such architecture innovations from both industry and academia that all share the same goal: to make FPGAs better at DL.

\section{FPGA Architecture \& Opportunities for DL}
\label{sec:fpga_arch}

In this section, we briefly describe the key building blocks of an FPGA architecture and highlight the opportunities for optimizing these different components for DL compute.
For a more comprehensive survey on the design principles and evolution of FPGA architecture, we refer the reader to~\cite{boutros2021fpga}.

\subsection{Programmable Logic \& Routing}
\label{sec:logic}

\begin{figure}[!t]
\centering
\includegraphics[width=0.98\linewidth]{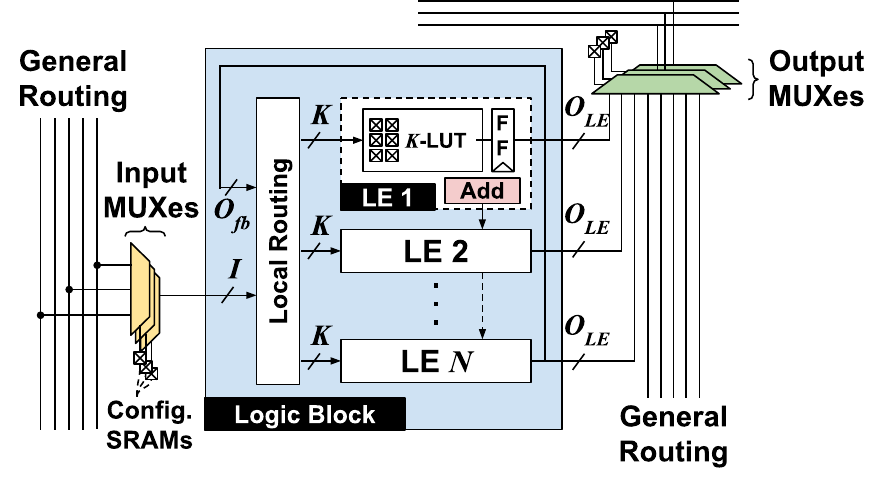}
\caption{Logic block (LB) and routing architecture. An LB consists of $N$ Logic Elements (LEs) and local interconnect. SRAM-controlled programmable routing MUXes connect general routing wires to each other and to LB inputs/outputs.}
\label{fig:logic_block}
\end{figure}

The programmable logic blocks (LBs) are the most abundant resource in an FPGA. 
An LB is a group of $N$ logic elements (LEs) in addition to local routing, generally built with programmable multiplexers, that allow the LB inputs to connect to different LEs or feed the LE outputs back to their inputs, as illustrated in Fig.~\ref{fig:logic_block}.
In their simplest form, each LE combines an SRAM-based $K$-input look-up table (LUT) with a bypassable flip-flop (FF) and can implement any optionally-registered $K$-input Boolean logic function.
The LEs in many modern FPGA architectures can also be fractured to implement two logic functions that use at most $K-1$ inputs each and together do not use more distinct inputs than the local routing provides to a single LE.
They also include dedicated circuitry (the pink box in Fig.~\ref{fig:logic_block}) to efficiently implement adders, which are abundant in many FPGA designs~\cite{murray2020optimizing} and very common in DL accelerators. 
Most commercial FPGAs from both AMD and Intel implement LBs of eight to ten 6-input LEs (i.e. $N=8-10, K=6$).
A key distinction is that each LE in Intel FPGAs includes dedicated circuitry to implement two bits of addition, while each AMD LE can only implement a single bit of addition. 
The LBs (and other FPGA fabric components and IOs) are surrounded by  programmable routing that can flexibly connect between various blocks.
This programmable routing consists of SRAM-controlled multiplexers to connect block outputs and routing wires to different routing wires (green MUXes in Fig.~\ref{fig:logic_block}) and routing wires to block inputs (yellow MUXes in Fig.~\ref{fig:logic_block}). 
These multiplexers consume a large fraction of an FPGA's die area; they constitute more than 50\% of the area of a logic \emph{tile} (i.e. LB and its programmable routing)~\cite{boutros2023into}. As adding more inputs and outputs to an FPGA LB or hard block implies more routing multiplexers, architecture changes that increase the number of inputs/output to/from a block require careful consideration of the functionality gain vs. the area cost.

The programmable logic and routing are the key to the FPGA's bit-level programmability and allow it to implement any functionality by setting LUT and routing configuration SRAMs (shown in Fig.~\ref{fig:logic_block}) accordingly.
For DL, custom low-precision MAC units are typically implemented using the LUTs, FF, and dedicated adder circuitry in the FPGA's LEs.
For example, the Microsoft Brainwave FPGA-based DL accelerator implemented custom 8-bit floating-point (\texttt{msfp8}) compute units in the FPGA's programmable logic. This custom floating point format, which emphasizes dynamic range over precision, has 2.9$\times$ higher MAC density than traditional \texttt{int8} compute units~\cite{chung2017accelerating} and yields inference accuracy comparable to 32-bit floating point (\texttt{fp32}). 
When somewhat reduced accuracy is tolerable, low precision binary MACs which are realized as XNOR and population count (popcount) operations~\cite{zhang2021fracbnn} can also be used in DL models. 
This results in very small and efficient compute units when implemented on an FPGA. 
In other devices, the compute units are pre-fabricated to implement conventional, higher precision MACs and word-wide logic operations and hence the full efficiency gains of such extremely low precisions are not realized. 
While narrow bitwidth operations are already a strength of FPGAs, current FPGA LBs were architected before the rise of DL and its high demand for low-precision MACs, raising the question of whether DL-targeted LB changes could further improve their MAC efficiency.

\noindent
\emph{\textbf{\underline{Opportunity 1:}} Enhancing logic block architecture to implement more efficient narrow-bitwidth multiplication and addition operations can result in significant gains for low-precision DL acceleration.}

\subsection{Digital Signal Processing Blocks}
\label{sec:dsp}
Since DL workloads are dominated by MAC operations, digital signal processing (DSP) blocks are crucial when implementing FPGA-based DL accelerators.
DSP blocks are \emph{hard} (ASIC-style) blocks embedded in the FPGA fabric that implement multipliers and adders.
However, they are designed with some programmability to increase their usability in various FPGA designs while still maintaining their ASIC-like efficiency.
For example, DSP blocks in the Intel Arria 10 and Stratix 10 FPGA families have configurable circuitry to perform multiplications of different precisions (e.g. one \texttt{int27} or two \texttt{int18} multiplications) as well as optional pre-multiplication adders, a post-multiplication adder/accumulator, bypassable pipeline registers, and dedicated routing wires between DSP blocks in the same FPGA column.
These DSP blocks were originally designed for wireless communication and filtering applications, which remain a major market for FPGAs.
Therefore, they natively support numerical precisions that are widely used in this domain (e.g. 27$\times$27 and 18$\times$18 multiplications in Intel FPGAs, and 27$\times$18 multiplications in AMD FPGAs).
Although they can be used to implement MAC units in DL accelerators, these precisions are typically higher than what is commonly needed for DL inference resulting in underutilization of the DSP block features (or equivalently, wastage of silicon area as the DSP blocks are over-engineered for DL requirements).

\noindent
\emph{\textbf{\underline{Opportunity 2:}} Adding low-cost reconfiguration circuitry to enable fracturing the multipliers inside DSP blocks into more lower-precision multipliers (while maintaining backward compatibility) can enhance DL performance.}

Besides numerical precisions, these DSP blocks include other features beneficial to traditional FPGA applications like wireless communication. 
For example, the Intel Stratix 10 DSP block has a small constant coefficient bank and input cascading registers/interconnect for implementing efficient finite impulse response (FIR) filters.
These DSP features consume silicon area, but are less useful for DL computations; replacing them with more DL-focused features could improve DL efficiency at the cost of losing backward compatibility with traditional DSP blocks.

\noindent
\emph{\textbf{\underline{Opportunity 3:}} Replacing the DSP blocks originally designed for the wireless communications domain with new DL-targeted blocks can increase the compute density and efficiency of FPGAs for DL.}

\subsection{On-Chip Block Memories}
\label{sec:bram}
FPGAs also include a large number of on-chip SRAM memory blocks typically referred to as block RAMs (BRAMs).
These BRAMs (more than 10,000 blocks in modern FPGAs) are spatially distributed in columns throughout the FPGA fabric, as shown in Fig.~\ref{fig:taxonomy}.
The latest generations of Intel FPGAs contain a single type of BRAM with 20Kb capacity~\cite{agilexmem}, while AMD FPGAs have 36Kb BRAMs as well as larger but less common 288Kb RAMs (typically referred to as Ultra RAMs or URAMs)~\cite{versalmem}.
The core of these BRAMs is a fixed size SRAM array with the conventional peripheral circuitry for read/write operations such as row decoders, sense amplifiers, and write drivers. 
However, similarly to DSP blocks, these BRAMs include low-cost reconfiguration circuitry in their peripherals to enable implementing buffers with different width/depth (narrow and deep vs. shallow and wide buffers) and number of ports depending on the application needs ~\cite{yazdanshenas2017don}. 
For example, by setting a few configuration SRAM cells, a 20Kb BRAM can be used as a read-only memory (ROM), a single-port RAM, or a dual-port RAM with a 1b$\times$16K, 2b$\times$8K, 4b$\times$4K, 8b$\times$2K, 16b$\times$1K, 32b$\times$512, or 40b$\times$512 organization.
The FPGA BRAMs can all be accessed in parallel providing massive on-chip memory bandwidth (on the order of petabits per second) with only one or two cycles of access latency.
Additionally, they can be independently controlled and directly connected to the compute units by exploiting the flexibility of the FPGA's programmable routing.
These features are useful for low-latency massively parallel DL compute on FPGAs.
However, with the increasingly pressing need to bring compute even closer to data for higher efficiency, the thousands of on-chip memory blocks in an FPGA can potentially do more than just store data to be used by the compute units implemented in LBs and DSPs.

\noindent
\emph{\textbf{\underline{Opportunity 4:}} With advances in processing-in-memory technology, enhancing BRAMs with in-memory compute capabilities can provide thousands of parallel compute units on the FPGA at a relatively low cost.}

\subsection{Interposers}
\label{sec:interposers}

\begin{figure}[!t]
\centering
\includegraphics[width=\linewidth]{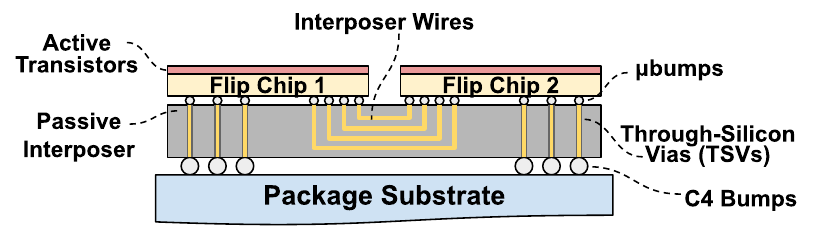}
\caption{Passive interposer technology for creating larger FPGA devices by integrating multiple smaller and higher yield chips in the same package.}
\label{fig:interposer}
\end{figure}

Since FPGAs are typically early adopters of a new process technology, creating large single-silicon-die FPGAs results in poor yield (due to manufacturing defects) especially early in the process life cycle.
To face this challenge, many recent FPGAs use passive interposer technology to integrate multiple (smaller) silicon dice in the same package.
This not only improves yield but also enables agile hardware development by combining FPGA fabrics with pre-fabricated \emph{chiplets} that implement different functionalities and (possibly) use different process technologies into a complete system-in-package.
As illustrated in Fig.~\ref{fig:interposer}, an interposer is a silicon die with conventional metal layers but has no active transistors implemented on it (thus the name \emph{passive interposer}).
The top metal layer of the interposer die can connect to the top metal layer of multiple dice flipped on top of it through densely packed (typically tens of $\mu m$ pitch) solder balls known as \emph{microbumps}~\cite{lau2022recent}, providing a high density of routing tracks between different chips in the same package.
AMD FPGAs have been using this technology starting from their 28nm 7-series family to integrate multiple FPGA dice which are presented to users as a single large FPGA with multiple super logic regions (SLRs)~\cite{ravishankar2018placement}.
Intel FPGAs also use a similar technology, known as the embedded multi-die interconnect bridge (EMIB)~\cite{mahajan2016embedded}, to integrate an FPGA die with multiple transceiver or high-bandwidth memory (HBM) chiplets starting from their 14nm Stratix 10 family~\cite{greenhill20173}.
These technologies enable the creation of many device variations for different markets depending on the application-specific ASIC chips integrated with the FPGA in the same package.
Even with the rapid change in state-of-the-art DL models, massively parallel MAC operations are a core component of almost all models and hence can be potentially offloaded to a highly-efficient ASIC side chiplet.
In this case, the FPGA in the same package can provide the needed flexibility for any DL model changes and diverse IOs to the rest of the system.

\noindent
\emph{\textbf{\underline{Opportunity 5:}} Integrating FPGAs and DL-specialized ASICs using advanced package integration technologies can combine the best of both worlds: FPGA flexibility for bespoke parts of the system and ASIC efficiency for common functionality.}

\subsection{Networks-on-Chip and Embedded Accelerators}
\label{sec:nocs}
More recently, new \emph{beyond-FPGA} reconfigurable acceleration devices (RADs) have emerged. 
An example is the AMD Versal architecture, which combines an FPGA fabric with general-purpose processor cores and an array of software-programmable vector processors in a single device~\cite{gaide2019xilinx}.
All these components are connected via a packet-switched network-on-chip (NoC) for efficient system-level communication~\cite{swarbrick2019network}. 
The NoC enables faster and easier integration of systems combining various \emph{soft} design IPs implemented on the programmable FPGA fabric along with \emph{hard} coarse-grained application-specific embedded accelerators.
The AMD Versal architecture is a single design point from a broad space of potential novel reconfigurable computing devices that could benefit DL acceleration.

\noindent
\emph{\textbf{\underline{Opportunity 6:}} Exploring the design space of new DL-targeted RADs that combine the unique features of FPGAs with more efficient coarse-grained DL accelerator cores.}

\vspace{0.2cm}
For the remainder of this article, we review recent proposals from both academia and industry to enhance FPGA architecture for DL, capitalizing on the opportunities that we highlighted in this section.
Before that, we first explain the commonly used methodology for exploring and evaluating new FPGA architectures quantitatively.  
\section{FPGA Architecture Exploration}
\label{sec:arch_exploration}

\subsection{Tools and Benchmarks}

\begin{figure}[!t]
\centering
\includegraphics[width=\linewidth]{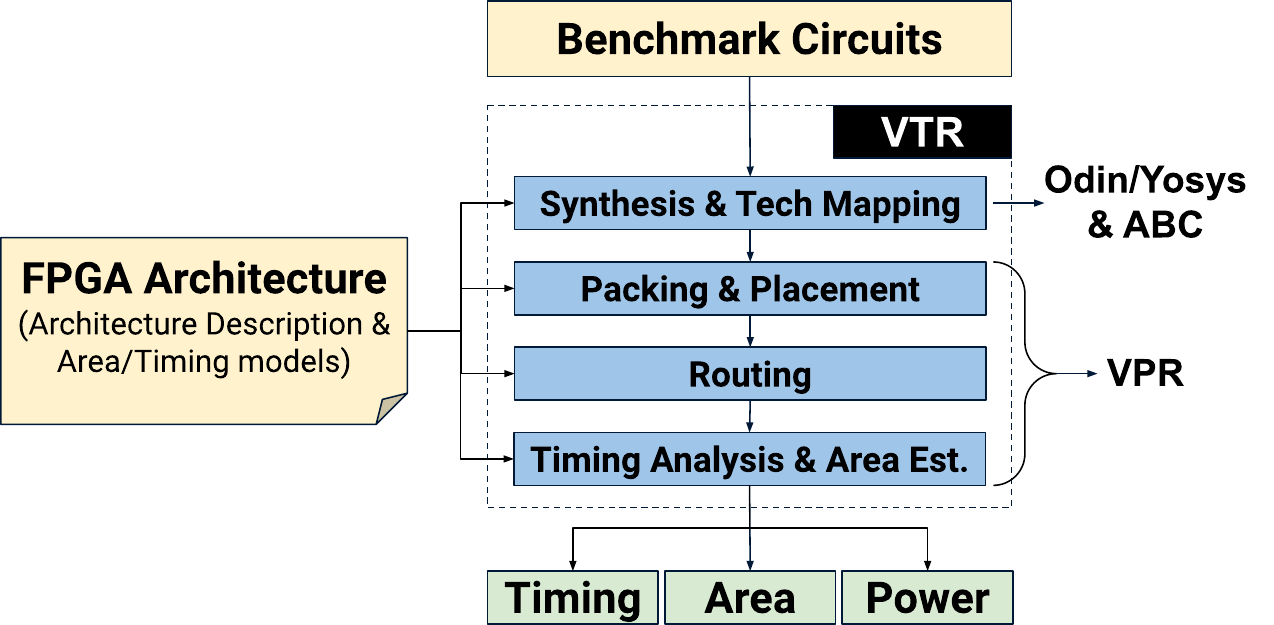}
\caption{Key ingredients for FPGA architecture exploration: benchmark circuits, architecture description, and a retargettable CAD flow.}
\label{fig:arch_eval_flow}
\end{figure}

Fig.~\ref{fig:arch_eval_flow} shows the flow typically used to evaluate FPGA architecture modifications.
At the core of this flow is a \emph{retargettable} FPGA CAD tool that can flexibly synthesize, place and route a set of benchmark designs on a wide range of input FPGA architectures. 
Architects can then evaluate different candidate architectures by comparing the the timing, area, and power metrics reported by the CAD tools.
Verilog-to-routing (VTR) is an open-source flow that is widely used for FPGA architecture and CAD research~\cite{murray2020vtr}. 
It combines several tools such as ODIN~\cite{jamieson2010odin} or Yosys~\cite{wolf2013yosys} for Verilog synthesis, ABC~\cite{brayton2010abc} for logic optimization and technology mapping, VPR for packing, placement and routing, and Tatum~\cite{murray2018tatum} for static timing analysis.
VTR takes as an input an XML-based FPGA architecture description file which specifies the high-level organization of an FPGA (e.g. number and type of blocks, distribution of wire segment lengths, size of logic clusters and logic elements), its micro-architectural details (e.g. DSP and BRAM modes of operation, hard arithmetic in logic blocks, switch block patterns), as well as transistor-level circuit implementation parameters (e.g. switch/wire delays and areas).
Tools such as COFFE~\cite{yazdanshenas2019coffe} automate the transistor-level design and modeling of FPGA circuitry and generate the delay and area of different components to be included in the VTR FPGA architecture description file. 

Optimizing an FPGA architecture also requires benchmark designs that cover a variety of use cases in the target application domains. 
Typically, FPGA vendors have carefully curated benchmark suites comprising proprietary designs representative of different customer use cases.
There are also several academic open-source benchmark suites such as the classic MCNC20, the VTR \cite{murray2020vtr}, and the Titan23 \cite{murray2013titan} suites which are commonly used in academic FPGA architecture and CAD research. 
While early academic FPGA research used the MCNC circuits, they are now too small (thousands of logic primitives) and simple (only IOs and logic) to represent modern FPGA applications. 
The VTR and particularly the Titan suites are larger and more complex, making them more representative.
However, none of these benchmark suites contain any FPGA designs representative of the DL domain.
The Koios benchmark suite~\cite{arora2023koios} was introduced to address this gap.
It contains 40 DL circuits that capture a wide variety of sizes, implementation styles, target neural networks, numerical precisions, and circuit properties.
It also introduced a methodology for generating synthetic or \emph{proxy} circuits that have similar characteristics as various real DL circuits.
The Koios benchmarks are open-sourced and integrated into the VTR flow, enabling the exploration of new FPGA architectures optimized specifically for DL.

\subsection{Methodology}

In this subsection, we explain the general methodology for evaluating new FPGA architecture ideas using the tools and benchmarks that we introduced in the previous subsection.
A similar methodology is used for evaluating the gains and cost of most of the proposed FPGA fabric architecture enhancements discussed in the rest of this article.

A common FPGA architecture enhancement for a specific target domain is to introduce a new type of hard block to the FPGA fabric (or change an existing one) to efficiently implement common functionalities in application designs from this domain.
As an example, for the DL target domain, an FPGA architect might evaluate adding hard convolution engines to the FPGA fabric.
This involves many design trade-offs and questions including: how much of the FPGA die area should be dedicated to these convolution blocks? How flexible should they be? Do they implement only convolutions or can be re-configured to implement other operations and used by a broader set of applications? What impact does their addition have on the demand for  programmable routing? How much do they improve the overall target application performance and at what cost to other application domains? 

To answer these questions, the architect first writes an \textbf{RTL implementation} for their new proposed hard block (a convolution engine in our example).
This implementation describes the cycle-accurate functionality of the block as well as its different reconfigurable modes of operation.
Then, they perform the \textbf{circuit-level evaluation} using a tool like COFFE. FPGA circuitry consists of both standard cell (ASIC) components and full custom (hand-optimized transistors and layouts) components, and COFFE models and evaluates both types. The functionality of the hard block is implemented using standard cell ASIC EDA tools, while the interface to the programmable routing uses automated full-custom design and SPICE modeling. 
The outcomes of this step are the area, timing and power models of the proposed hard block.
These models are then plugged into an FPGA CAD flow (such as VTR) to perform the \textbf{architecture-level evaluation} by mapping a set of representative benchmark circuits (such as Koios) to the FPGA architecture including the proposed new hard block.
This mapping can be done by modifying the benchmarks to directly instantiate an instance of the new hard block or by extending the synthesis tools to automatically extract circuit netlist components and map them to the new hard block.
This last step evaluates the resource utilization, timing, and routability of the benchmarks of interest on the proposed FPGA architecture.
Enhancements to the programmable logic blocks and BRAMs can be evaluated using the same general methodology, except that the core of these blocks is also custom designed and laid out instead of being implemented with ASIC standard cells.

\subsection{Taxonomy of FPGA Architecture Enhancements for DL}

\begin{figure}[!t]
\centering
\includegraphics[width=0.9\linewidth]{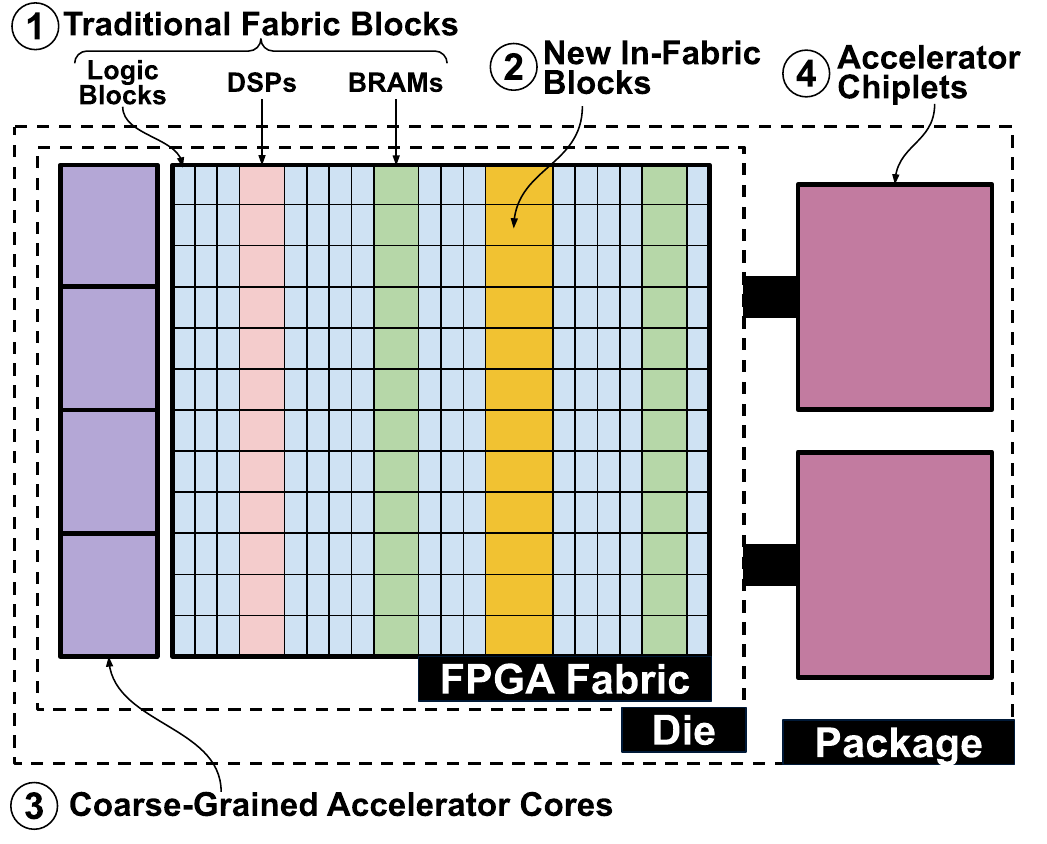}
\caption{Taxonomy of FPGA architecture enhancements for DL.}
\label{fig:taxonomy}
\end{figure}

In the rest of this article, we present several DL-targeted FPGA architecture enhancements from both academic research and industry.
Fig.~\ref{fig:taxonomy} illustrates the taxonomy of these proposals which include \circled{1} enhancing existing conventional FPGA fabric blocks (e.g. LBs, DSPs, BRAMs), \circled{2} introducing new DL-specific in-fabric hard blocks (e.g. tensor blocks), \circled{3} tightly integrating coarse-grained DL accelerators on the same die with an FPGA fabric (e.g. AMD AI engines), as well as \circled{4} integrating FPGAs and other DL chiplets in the same package.
\section{Enhancing Existing FPGA Fabric Blocks}
\label{sec:existing_blocks}

\subsection{Logic Blocks}

As discussed in Section~\ref{sec:logic}, many prior works have shown that various DL models can be quantized down to lower precisions with little to no accuracy loss during inference~\cite{wu2020integer}.
Narrow integer MAC operations (e.g. \texttt{int8}, \texttt{int4}) are now natively supported in many commercial DL accelerators~\cite{abts2020think, anderson2021first}.
In addition, new low-precision floating point formats are being standardized (e.g. \texttt{fp8}~\cite{micikevicius2022fp8}) and are expected to be supported in the next generation of compute platforms for DL.
FPGAs offer unique flexibility to implement any desired numerical precision directly in hardware using their fine-grained programmable LBs and since the logic usage of a multiplier grows quadratically with its precision, large hardware savings are possible by cutting precision to the minimum.
LBs are the most abundant resource in a conventional FPGA fabric, so architecture changes that increase their efficiency for implementing low-precision multipliers will have high impact in DL applications.

\begin{figure}[!t]
\centering
\includegraphics[width=\linewidth]{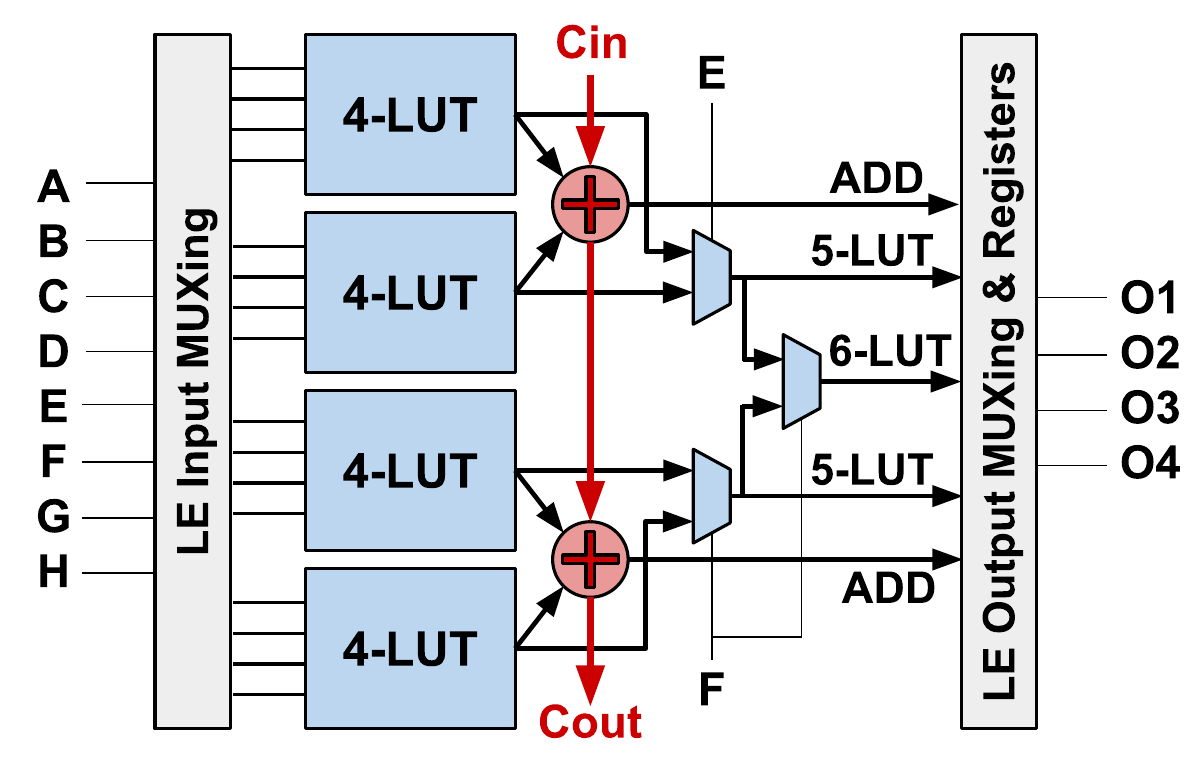}
\caption{Architecture of a logic element similar to that of Intel Stratix 10 and Agilex. It can operate as four 4-LUTs followed by two additions, two 5-LUTs, or one 6-LUT.}
\label{fig:logic_element}
\end{figure}

\begin{figure}[!t]
\centering
\includegraphics[width=\linewidth]{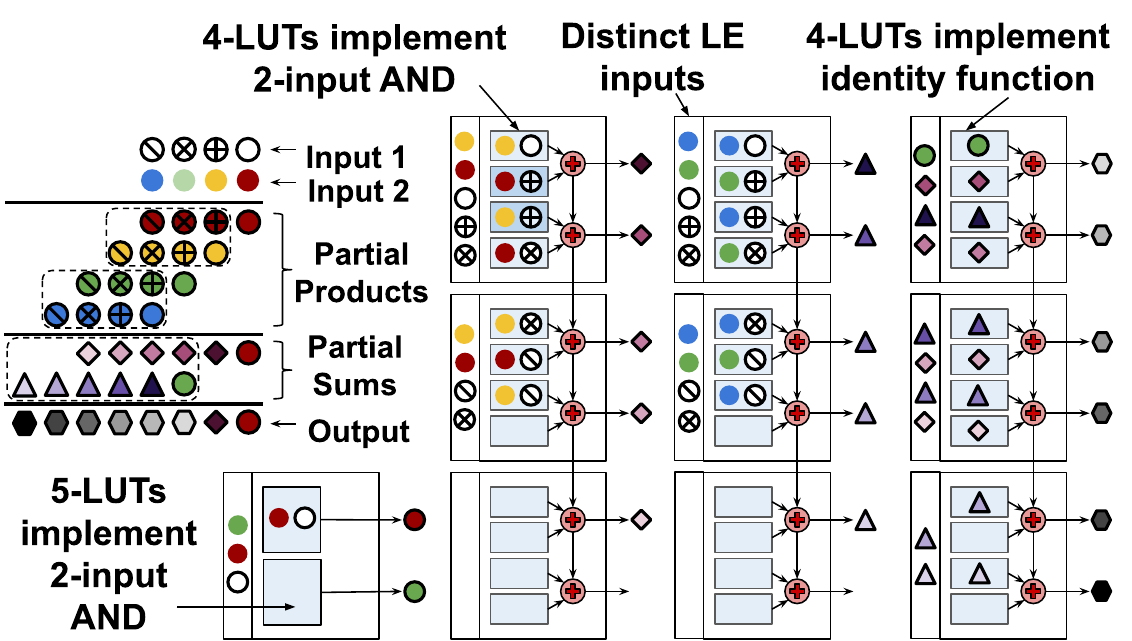}
\caption{Mapping of 4-bit multiplication to conventional logic elements. The LUTs are significantly underutilized: 4/5-LUTs are used to implement 2-input ANDs or the identity function (i.e.~pass-through) to access the adders.}
\label{fig:mult_mapping}
\end{figure}

Fig.~\ref{fig:logic_element} shows the internal architecture of a modern FPGA LE, similar to that in the Intel Stratix 10 and Agilex FPGAs.
It has 8 distinct inputs and 4 optionally-registered outputs, as well as two chained hard adders that are fed by four 4-input LUTs (4-LUTs).
Therefore, at a high-level, each LE can implement four 4-input logic functions followed by 2-bits of addition, two 5-input logic functions, or one 6-input logic function as long as no more than 8 distinct inputs are needed.
Fig.~\ref{fig:mult_mapping} illustrates how a 4-bit multiplication is mapped to this LE architecture as an example, where bits of the two multiplication operands are represented by different shapes and colors.
The first step of multiplication is performing an AND between each bit of one operand and all bits of the other operand to generate partial products, as illustrated by combining the color and shape of the bits in Fig.~\ref{fig:mult_mapping}.
These partial products are then reduced over one or multiple stages of addition to produce the final multiplication result.
Thus, multiplications can be fundamentally viewed as bit-level ANDs followed by adder trees (usually referred to as \emph{compressor trees}).
In an LE implementation, the 2-input ANDs are mapped to the 4-LUTs followed by adders to realize the first level of reduction.
Then, other LEs are used only for the adders (i.e. the 4-LUTs implement identity functions) to perform the subsequent reductions until the final result is produced.
This highlights a major source of inefficiency: the LUTs are significantly underutilized.
Many of the used LUTs in Fig.~\ref{fig:mult_mapping} are just pass-throughs to access the adders, and even the LUTs that implement partial products perform a 2-input AND function, wasting half of the functionality of a 4-LUT.

\begin{figure}[!t]
\centering
\includegraphics[width=\linewidth]{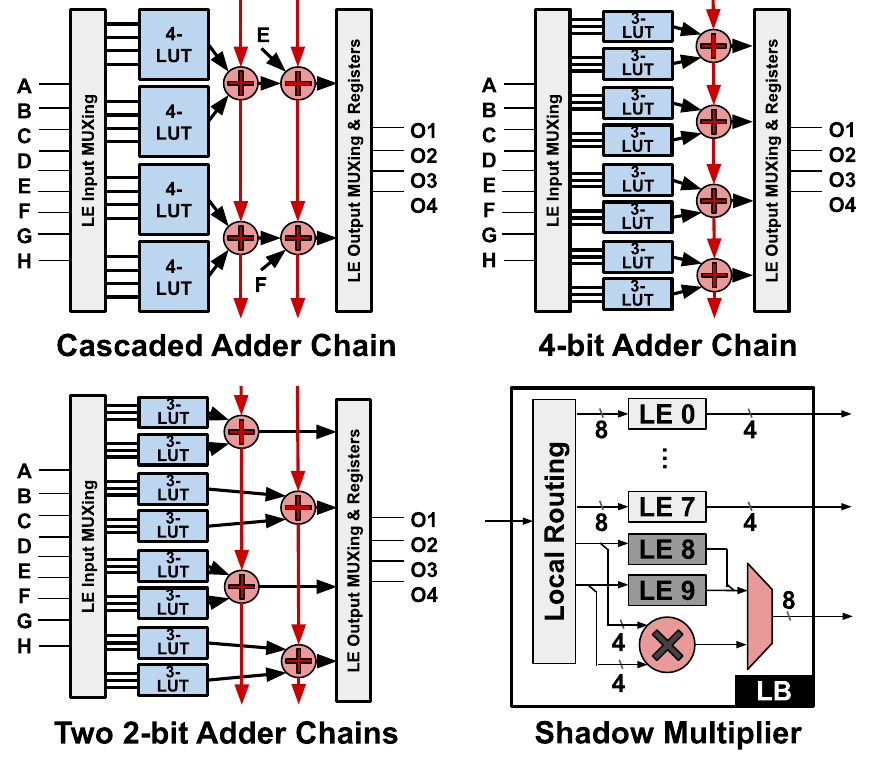}
\caption{Four architectural modifications to the FPGA LEs and LBs for increasing the density of low precision MAC in soft logic.}
\label{fig:logic_enhancements}
\end{figure}

The authors of~\cite{boutros2019math,eldafrawy2020fpga} highlight these inefficiencies and propose four architectural modifications (summarized in Fig.~\ref{fig:logic_enhancements}) to address them at both the LE and LB levels. 
The first proposal adds another \textbf{cascaded adder chain} fed by the two sum outputs of the existing chain and two independent inputs.
This can efficiently implement compressor trees by obviating the need for a second level of LEs used only as adders.

The second proposal implements a single \textbf{4-bit adder chain} by adding circuitry to allow further fracturing of each 4-LUT into two 3-LUTs. 
Fracturing a 6-LUT all the way down to 3-LUTs generates 8 signals which can feed two inputs into each of four adders, and a 3-LUT can still implement the 2-input AND gate function needed in multiplier partial products.
This results in a higher density of adders in general at the cost of sacrificing the ability to map 4-input logic functions feeding adders in the same LE, which does not occur in multipliers and is expected not to be very common in other non-DL FPGA designs. 
While one might consider continuing this process, fracturing down to 2-LUTs feeding 8 chained adders per LE, this would exceed the 8 inputs the programmable routing provides to an LE, and adding additional inputs from the programmable routing is costly in terms of area~\cite{boutros2021fpga}. 
Another variation on this idea arranges the four adders into \textbf{two 2-bit chains} per LE instead of a single 4-bit chain, as shown in Fig.~\ref{fig:logic_enhancements}.
This is different from the cascaded adder chain in the first proposal, since all the adders in both chains are fed directly by the LUTs.

The fourth idea modifies the LB architecture by adding a low-precision hard multiplier in some or all of the FPGA LBs. 
They are referred to as \textbf{shadow multipliers} as, when used, they steal the input and output ports of some of the LEs; this makes those LEs unusable but avoids adding expensive input and output ports to the programmable routing. 
The shadow multipliers from different LBs can also be combined to implement higher-precision multiplications using the programmable routing and some LE-based glue logic.

These four ideas vary in their area costs and performance gains.
For example, the two 2-bit adder chains proposal results in 1.5$\times$ denser matrix multiplications, while being 10\% faster and also benefiting other non-DL benchmarks.
These gains come at a modest cost of only 3\% increase in die area compared to a Stratix-10-like baseline fabric.
On the other hand, adding a 9-bit shadow multiplier to each LB results in 2.4$\times$ denser and 17\% faster matrix multiplications, at the cost of a 15\% increase in die area.
A patent filed by Intel~\cite{yazdanshenas2021efficient} further enhanced the cascaded adder chains proposal to achieve denser MAC mappings, but is not yet adopted in commercial FPGA architectures.

While~\cite{boutros2019math, eldafrawy2020fpga} focused on adding more full adders to LEs for denser arithmetic, both MAC and pop-count operations in low-precision and binarized DL models typically require the addition of more than 3 bits and can benefit from generalized parallel adders or \emph{compressors}.
A full adder is a simple compressor that takes 3 bits as inputs ($A, B, C_{in}$) and \emph{compresses} them into 2 bits (a sum $S$ and a higher significance carry $C_{out}$).
Therefore, a full adder is typically referred to as a $C3:11$ compressor (3 inputs $\rightarrow$ 1 same significance + 1 higher significance outputs).
This concept can be generalized to any number of input bits, where the compressor output is simply a count of the number of ones in the input bits.
The authors of~\cite{rasoulinezhad2020luxor} analyzed a variety of microbenchmarks and found that more than 35\% of the compressors in these designs are $C6:111$ compressors.
A $C6:111$ can be viewed as three 6-input logic functions (one for each output bit) and thus can be mapped to 3 LEs.
One of these three logic functions is a simple 6-input XOR.
In~\cite{rasoulinezhad2020luxor}, the authors evaluated adding a hardened 6-input XOR gate to a typical recent LE architecture (similar to that in Fig.~\ref{fig:logic_element}).
Since all three logic functions share the same 6 inputs and the LE has up to 4 outputs, the added XOR gate enables a single LE to implement two of the three logic functions in a $C6:111$ compressor.
This results in up to 36\% denser compressor implementations at the cost of less than 0.5\% increase in die area. 

Kim et al.~\cite{kim2018fpga} also proposed two architectural modifications to the hard adder chains in LBs for enhancing the efficiency of popcount operations in binarized DL models.
The first proposal adds a new popcount adder chain that propagates the sum bits on the dedicated chain interconnect and produces the carry out bits to the outputs of the LEs (which differs from the conventional adder chain that propagates carry bits and produces sum bits).
The second proposal further optimizes popcount implementation by adding another full adder that can sum the two carry out bits of the two popcount adders in an LE.
These two architectural changes reduce the logic utilization of popcount operations of different widths by 23-44\% and 36-40\% at the cost of only 1.9\% and 2.4\% increase in the LB's silicon footprint, respectively.

\subsection{DSPs}
\label{sec:dsp_enhancements}

Along the same vein of increasing low-precision MAC efficiency on FPGAs, both academic research and FPGA vendors have investigated adding native support for low precisions in conventional DSP blocks.
As discussed in Section~\ref{sec:dsp}, filtering and wireless communication applications were historically the key drivers of DSP block architecture decisions.
Therefore, DSP blocks in commercial FPGAs until the 14nm process generation from both Intel (Stratix 10) and AMD (Ultrascale+) had native support for numerical precisions suitable for wireless communication applications.
In 2013, Intel added native support for single-precision floating-point (\texttt{fp32}) in the DSPs of their Arria 10 (and later Stratix 10) devices to enhance their efficiency for high-performance computing.
The rapid growth in the DL domain motivated the work in~\cite{boutros2018embracing}, which was the first to investigate DSP micro-architecture optimizations for low-precision DL.

\begin{figure}[!t]
\centering
\includegraphics[width=\linewidth]{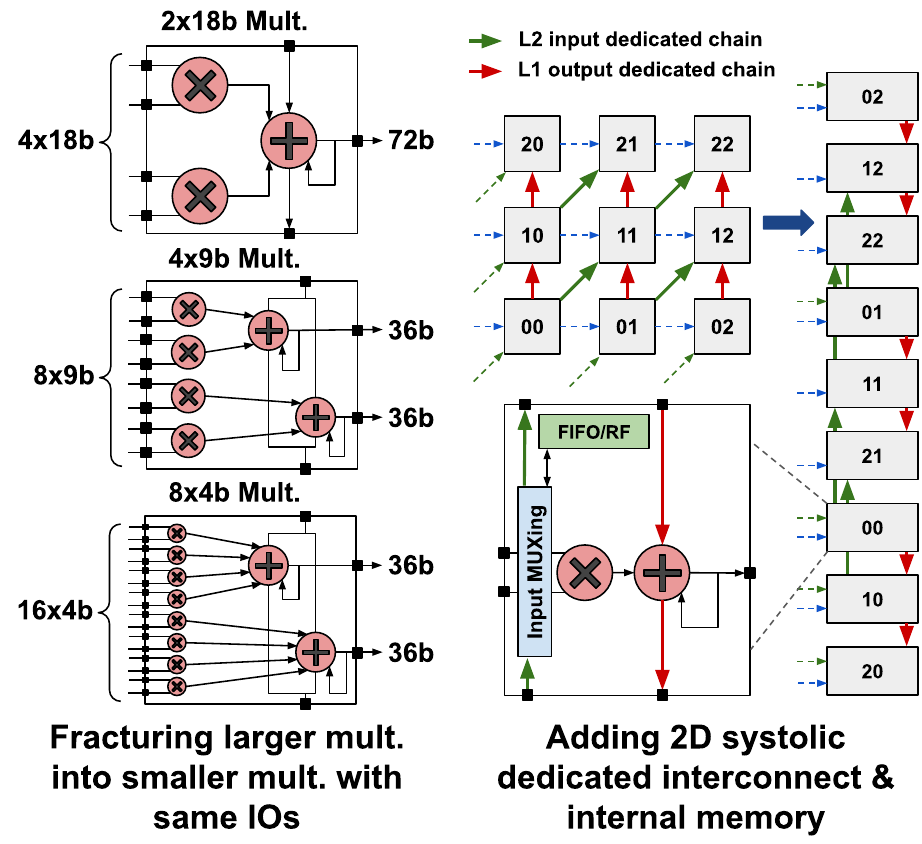}
\caption{Enhancements to the FPGA DSP blocks for DL such as: (1) fracturing larger multipliers into smaller ones while keeping the same interface to the programmable routing (left), and (2) adding more dedicated interconnect between DSPs in a column for efficient 2D systolic array implementation and integrating an internal FIFO or register file for efficient data reuse near compute (right).}
\label{fig:dsp_enhancements}
\end{figure}

This work enhanced an Arria-10-like DSP block that can implement one \texttt{int27} or two \texttt{int18} multiplications to also natively support four \texttt{int9} and eight \texttt{int4} multiply and MAC operations at a low area cost, as illustrated on the left side of Fig.~\ref{fig:dsp_enhancements}.
This was achieved by balancing the addition of new small 4-bit multiplier arrays and low-cost circuitry that enables the fracturing of existing multiplier arrays into multiple independent sub-arrays.
In addition, the chain reduction and accumulation was split into two lanes as shown in Fig.~\ref{fig:dsp_enhancements} to minimize the area and delay cost of supporting the MAC mode for these precisions.
The design of this new enhanced DSP block was guided by three key design principles: (1) Ensure \textbf{backward compatibility} such that the DSP blocks are still efficiently usable for non-DL applications, (2) Have \textbf{minimal effect on DSP block area footprint and operating frequency} to minimize the negative impact on other applications that do not benefit from the added modes of operation, and (3) Keep the \textbf{same number of input/output ports} to/from the DSP block to avoid both the expensive area cost of additional interfaces to the programmable routing and the creation of routing hot spots in the proximity of these blocks.

The enhanced DSP block from~\cite{boutros2018embracing} increased the area of the DSP block by 12\% which corresponds to only a 0.6\% increase in the overall die area of DSP-rich devices, with no effect on its operating frequency.
When used in several DL accelerator designs, the new DSP blocks enhanced performance by 1.3$\times$ and 1.6$\times$ while reducing resource utilization by 15\% and 30\% for \texttt{int9} and \texttt{int4} precisions, respectively.
Subsequent commercial FPGA architectures from both Intel (Agilex) and Xilinx (Versal) added similar native support for four and three \texttt{int8/int9} multiplications per DSP block, respectively.  

Conventional DSP blocks also have dedicated wires that can pass the inputs/outputs of one DSP block to the next block in the same column.
This was originally designed to help implement more efficient 1D systolic arrays for finite impulse response (FIR) filters in wireless communication applications.
However, for the DL domain, efficient matrix-matrix multiplication and convolution operations can be implemented as 2D systolic arrays.
Therefore, Rasoulinezhad et al.~\cite{rasoulinezhad2019pir} explored \textbf{adding a special pattern of dedicated interconnect between DSP blocks} that can efficiently map 2D systolic arrays to a 1D column of DSP blocks without using the general programmable routing, as shown on the right side of Fig.~\ref{fig:dsp_enhancements}.
They also proposed \textbf{integrating a small memory inside the DSP block} (register file or FIFO) to enhance energy efficiency by storing data very close to compute.
This enables reusing the same set of operands across many computations (which is common in many DL compute kernels) without the need to read and transport it from distributed LUT-based memories or BRAMs to DSP blocks.
Their PIR-DSP block significantly reduced energy consumption by 70\%, 82\%, and 87\% (on average across several neural network implementations) compared to a baseline Xilinx-like DSP block for \texttt{int9}, \texttt{int4}, and \texttt{int2} precisions, respectively.
These improvements come at the cost of 28\% increase in the DSP block area footprint.

\subsection{BRAMs}

In DL applications, the FPGA BRAMs are used as on-chip user-managed scratchpads to store computation operands (weights and activations) and results, feeding the compute units with data at a very high bandwidth due to their distributed nature. 
However, the separation of compute units (implemented using LBs and DSPs) from storage units (BRAMs) implies data movement to feed the compute units with input data and store the outputs back to the BRAMs. 
This uses a large amount of the FPGA's programmable routing, leading to routing hot spots between the memory and compute units and increased power consumption.
To address these challenges, several research efforts have proposed adding compute capabilities to the FPGA BRAMs by introducing lightweight bit-level PEs inside the BRAM itself to bring compute closer to the data.
This provides three major advantages: (1) It \textbf{increases the compute throughput} of the FPGA because a larger portion of the FPGA die area can now perform computation, (2) it \textbf{reduces the data movement} saving both energy and valuable programmable routing resources,
and (3) it \textbf{provides massive compute parallelism} as the large number of BRAM bitlines can operate as bit-serial SIMD lanes executing the same operation on all the bits of a memory word.
Similarly to the DSP block enhancements in Section~\ref{sec:dsp_enhancements}, the new compute-capable BRAMs should be functionally backward compatible, incur minimal performance loss for traditional usage, and avoid adding new input/output ports to/from the programmable routing to avoid area overheads for designs that do not use in-BRAM compute.

At a high level, enabling in-BRAM compute requires adding PEs that perform bit-serial computations on the outputs of the sense amplifiers inside the BRAM.
Two $N$-bit rows (i.e. wordlines) are read simultaneously from the RAM cell array, the PEs perform $N$ parallel binary operations between the corresponding bits of the two words, and the result is stored back to another row in the RAM array.
This read-modify-write operation happens completely inside the BRAM in a single clock cycle that is longer than the normal read/write period in a conventional BRAM.
In addition to the sense amplifier PEs, lightweight control logic (a finite state machine) may be required inside the BRAM to sequence these steps.
The specific rows to read, computation to perform in the PEs, and row to write constitute a compute instruction which is provided to the BRAM through the existing programmable routing ports.

\begin{figure}[!t]
\centering
\includegraphics[width=0.8\linewidth]{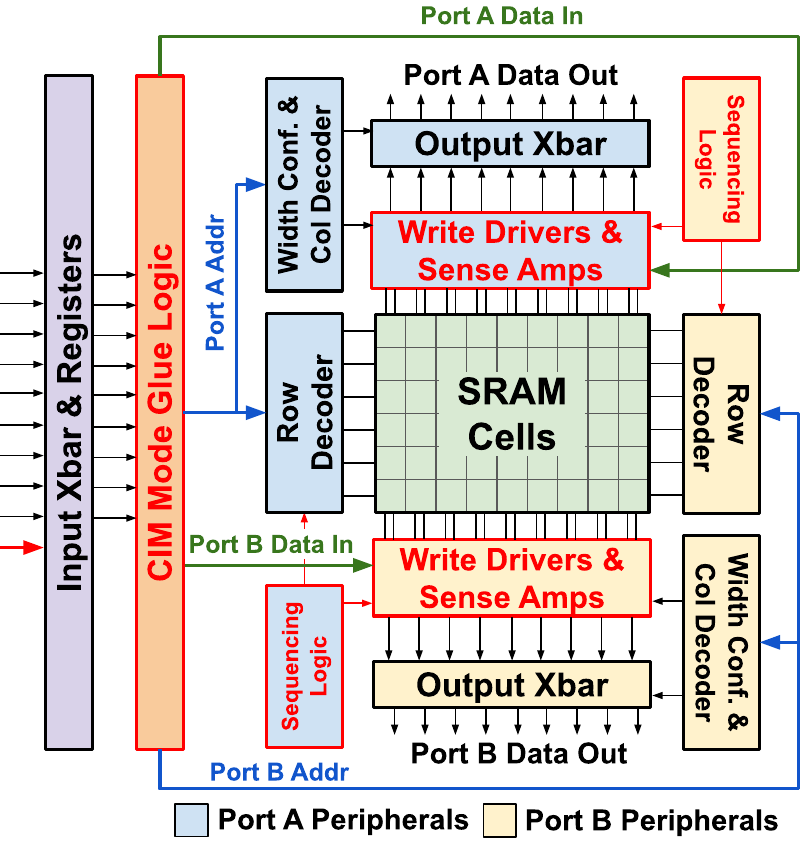}
\caption{FPGA BRAM internal architecture with the components changed or added for in-memory compute highlighted in red.}
\label{fig:bram_arch}
\end{figure}

Fig.~\ref{fig:bram_arch} shows a top-level diagram of an FPGA BRAM with the modified/added components to incorporate compute capabilities highlighted in red.
The dual-port memory cell array at the core of the BRAM remains unmodified.
In a conventional BRAM, the column decoder activates a subset of the bits in a row to be read by the sense amplifiers or written by the write drivers. 
For example, a 20Kb SRAM array (similar to that in the BRAMs of modern Intel FPGAs) is arranged as 128$\times$160-bit rows~\cite{tatsumura2016high}.
However, the maximum read/write width of the BRAM is 40 bits to limit the cost of the programmable routing interfaces.
Therefore, the BRAM block includes 40 sense amplifiers and 40 write drivers and a column decoder selects one 40-bit portion of the 160-bit row to be read/written.
To enable maximum parallelism for in-BRAM compute, additional sense amplifiers and write drivers as well as bit-level PEs are introduced to read/compute/write the full width of the array row.
The sequencing logic that controls the events of the read/write operations (wordline activation, precharge, sense amplifier enable, etc.) in the memory is also modified to support reading, computing, and writing in one (longer) cycle. 
One extra interface pin is added to the BRAM; when it is asserted the input data and addresses are treated as a compute-in-memory (CIM) instruction.
In this case the CIM mode glue logic decodes the instruction into low-level control signals to various BRAM internal components.

\begin{figure}
\centering
\begin{subfigure}{.45\linewidth}
  \centering
  \includegraphics[width=\textwidth]{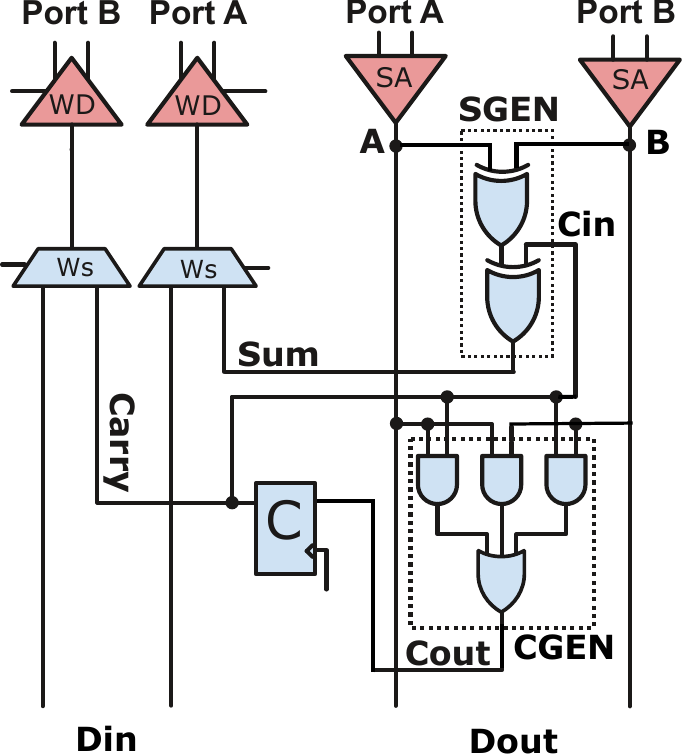}
  \caption{}
  \label{fig:cim_processing_element}
\end{subfigure}~~~
\begin{subfigure}{.5\linewidth}
  \centering
  \includegraphics[width=\textwidth]{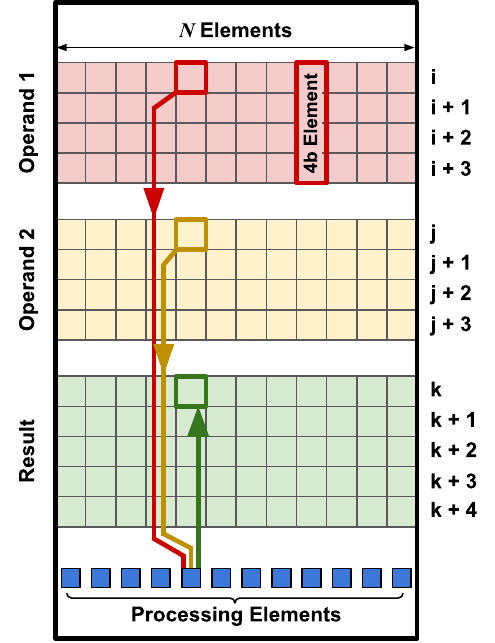}
  \caption{}
  \label{fig:cim_operation}
\end{subfigure}
\caption{(a) Compute-in-memory processing element circuitry for bit-serial addition. (b) In-BRAM compute operation example for elementwise addition of two $N$-element vectors with 4-bit operands. The input and result vectors are all stored in a transposed layout.}
\label{fig:cim}
\end{figure}

Fig.~\ref{fig:cim_processing_element} shows an example architecture of a CIM PE that can perform bit-serial addition. 
On the read path, \texttt{A} and \texttt{B} are the two operand bits read from two rows of the SRAM cell array by the sense amplifiers (\texttt{SA}) for the two SRAM array ports. 
The two XOR gates (\texttt{SGEN}) generate the sum bit (\texttt{Sum}) using the two operand bits (\texttt{A} and \texttt{B}) and the previous cycle's carry (\texttt{Cin}). 
Another set of gates (\texttt{CGEN}) are used to compute the carry bit, which is stored in the carry FF (\texttt{C}) for the next cycle computation. 
The read outputs \texttt{A} and \texttt{B} are also sent to \texttt{Dout} ports, which is the normal read path. 
On the write path, 2-input multiplexers (\texttt{Ws}) are added before the write drivers (\texttt{WD}) for the two ports. 
These multiplexers determine what to write to the SRAM cell array; 
the \texttt{Ws} multiplexers select between the sum/carry bit and the normal write path inputs (\texttt{Din}). 
The select lines of these multiplexers are driven by the CIM mode glue logic, depending on the mode setting and the instruction written to the BRAM input ports.

Fig.~\ref{fig:cim_operation} illustrates the operation of a compute-capable BRAM used to perform elementwise addition of two $N$-element vectors (operands 1 and 2), where each element is a 4-bit integer.
The vector elements are first stored in a transposed memory layout, where each of the elements of the first vector is stored in a different column over 4 rows ($i$ to~$i+3$) and elements of the second vector are stored in the same columns over 4 different rows ($j$ to~$j+3$). 
In one cycle, rows~$i$ and~$j$ are read, one on each port of the dual-port SRAM array.
Each PE receives two bits (one from row $i$ and one from row $j$) and computes the sum of the two bits and the carry from the previous cycle. 
The carry-out is stored in the carry FF in the PE, and the sum is written to row~$k$ using one port. 
This process is repeated for 4 cycles while incrementing the row addresses read from and written to. 
In the fifth cycle, the last carry bits stored in each PE are written to row $k+4$ using the second write port and the final result vector of the elementwise addition operation is now available in rows $k$ to~$k+4$.
More complex operations such as multiplication or reduction can be performed as a sequence of additions and memory copies.

There are several academic proposals to enhance FPGAs with in-BRAM compute capabilities; they make different design choices for the compute paradigm used (bit-serial vs. bit-parallel), supported operations in the added PEs, how to store data and intermediate results, and how to program/control the BRAMs to execute a sequence of operations~\cite{wang2021compute, arora2021compute, arora2022comefa, chen2023bramac, chen2023m4bram}.
The work by Wang et al.~\cite{wang2021compute} was the first to propose adding compute capabilities similar to that demonstrated for CPU caches~\cite{eckert2018neural} to FPGA BRAMs.
Their compute-capable BRAM (CCB) used a bit-serial addition PE; however, it uses only one port by activating two wordlines simultaneously to perform an analog AND operation on the bitlines.
This makes the PE slightly cheaper and frees up one of the two ports of the BRAM, enabling overlap of data loading and computation.
However, this technique is less robust, more sensitive to process variations, and requires lowering the wordline voltage (and therefore the operating frequency) to avoid corruption of the cell contents.
A DL accelerator designed for CCB achieves 1.25$\times$ and 3$\times$ higher performance compared to the Microsoft Brainwave accelerator~\cite{fowers2018configurable} for \texttt{int8} and \texttt{bfp8} precisions across RNN, GRU, and LSTM workloads, at the cost of only 1.8\% increase in the FPGA die area. 
Like CCB, Compute RAM~\cite{arora2021compute} performs analog AND operations on the bitlines and uses bit serial processing elements for addition, but introduces a small secondary memory array to store instructions inside the BRAM block.

CoMeFa~\cite{arora2022comefa} improves robustness over CCB by avoiding an analog AND on the bitlines; instead it exploits the dual-port nature of FPGA BRAMs to obtain two operands and uses the bit-serial addition PE from Fig.~\ref{fig:cim_processing_element}.
This technique also can achieve higher operating speeds, but it comes at the cost of using both BRAM ports during compute, and hence it cannot overlap loading and compute.
The CoMeFa architecture has both area- and delay-optimized variants; the delay-optimized version increases FPGA die area by 3.8\% and achieves a 2.5$\times$ performance improvement across a variety of DL workloads on a Microsoft-Brainwave-like accelerator architecture that uses in-BRAM compute.

Both CCB and CoMeFa followed the same bit-serial compute paradigm where the operands are laid out in memory in a transposed format.
In DL applications, one set of operands (the model weights) are fixed and therefore can be transposed offline and stored in the BRAMs.
However,~\cite{arora2022comefa} shows that implementing a data transformation unit to transpose the other set of operands (model activations) at runtime uses a significant amount of soft logic resources.
Chen~et~al.~\cite{chen2023bramac} instead proposed a compute-in-BRAM architecture for multiply accumulate (BRAMAC) that uses a mix of bit-serial and bit-parallel computations to reduce latency and enable the use of non-transposed activation values.
The PEs in BRAMAC are variable-precision adders that can take inputs from groups of bitlines to perform bit-parallel addition; multiplications are then implemented by serially accumulating addition results.
This significantly reduces the compute latency compared to the purely bit-serial approach: from $O(m^2)$ to $O(m)$ cycles for $m$-bit operands. 
However, it limits the possible numerical precisions to a pre-defined set supported by the architecture whereas the bit-serial approach can implement any precision.
BRAMAC also added a smaller secondary SRAM memory array with only a few wordlines inside the BRAM block.
In the compute mode, the operands are first copied internally (two 40-bit data words per cycle) to this secondary array, where the computations are performed.
This increases the BRAM compute mode frequency as it is faster to charge/discharge the much shorter bitlines of the secondary array; it also frees up both ports of the main memory array to be used for normal read/write operations while computations are performed in the secondary memory array.
Different variations of the BRAMAC architecture show performance improvements ranging from 1.3$\times$ to 2$\times$ for different CNNs running on an accelerator similar to Intel's DLA~\cite{abdelfattah2018dla} at the cost of a 3.4-6.8\% increase in FPGA die area.
M4BRAM~\cite{chen2023m4bram} augments the BRAMAC PE by adding duplication/shuffling logic to enable more efficient data reuse and adding support for mixed-precision operations in which the weights and activations have different bitwidths.
These enhancements improve the performance by 1.4$\times$ on average compared to BRAMAC.
\section{In-Fabric Tensor Blocks}
\label{sec:new_blocks}

\begin{figure}[!t]
\centering
\includegraphics[width=\linewidth]{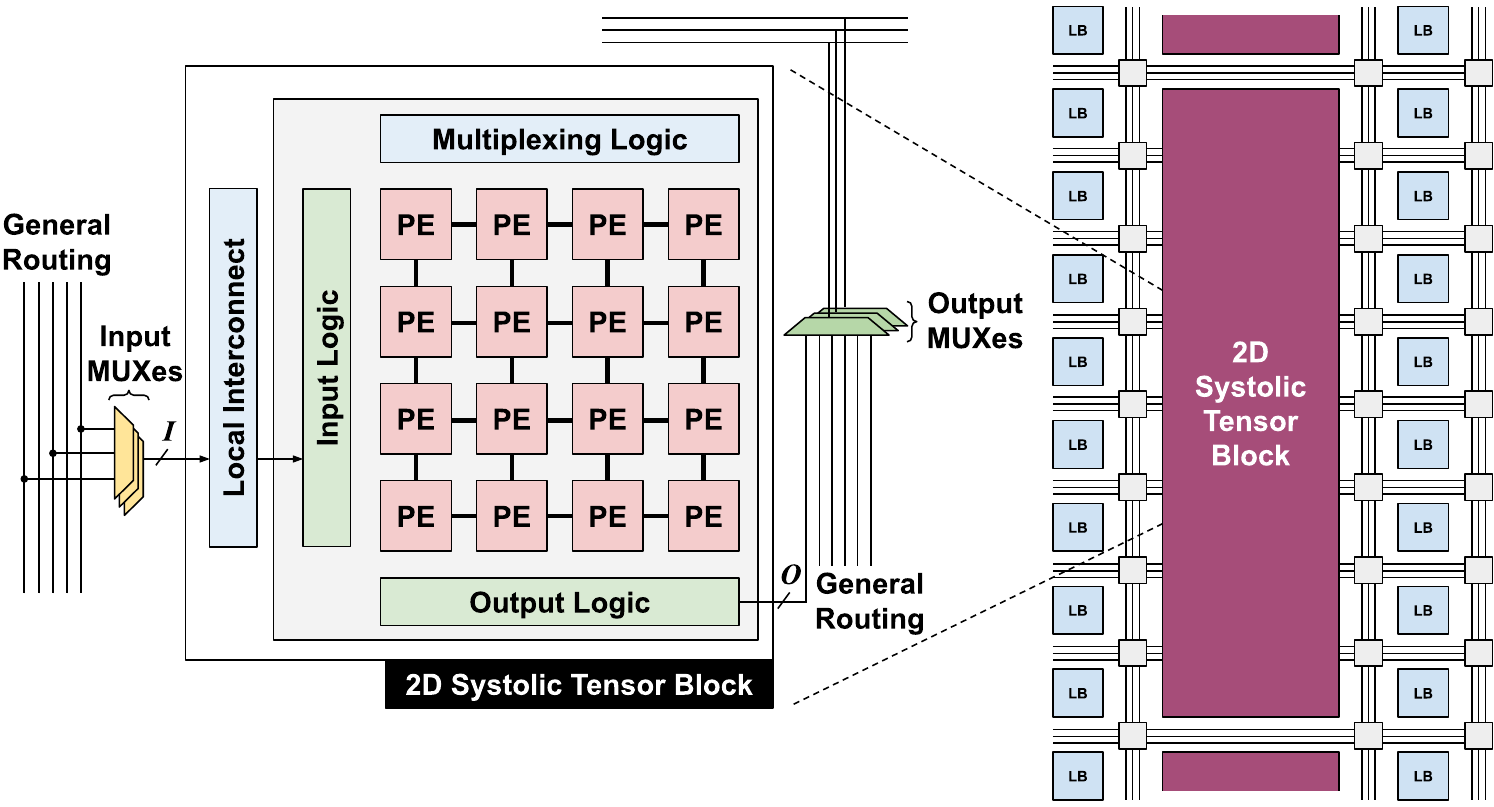}
\caption{In-fabric 2D systolic tensor block consisting of 16 PEs that can operate collectively in tensor mode or independently in scalar mode as configured by the multiplexing logic. Each tensor block is 3.5$\times$ wider than a LB and spans 8 rows in the FPGA grid.}
\label{fig:2d_systolic_tensor_block}
\end{figure}

Another line of work has investigated the integration of new hard blocks for tensor computations in the FPGA fabric to enhance DL inference efficiency.
Arora et al.~\cite{arora2021tensor, arora2022tensor} proposed adding 2D systolic tensor blocks to the FPGA fabric, as illustrated in Fig.~\ref{fig:2d_systolic_tensor_block}; these tensor blocks are in addition to (rather than a replacement for) the traditional DSP blocks in the fabric. 
These blocks contain 16 PEs, input logic for preparing data to be consumed by the PEs (e.g. delay registers for staggering inputs in 2D systolic processing), output logic for marshalling output data from different PEs, and multiplexing logic for configuring the block to operate in different modes.
The multiplexing logic allows the tensor block to operate in tensor or scalar modes.
In tensor mode, all the PEs are collectively calculating a matrix-matrix multiplication, matrix-vector multiplication, or matrix-matrix elementwise addition/subtraction/multiplication.
In scalar mode, each PE is calculating an independent multiply or MAC operation.
The mode of operation can be dynamically changed at runtime by appropriately setting control inputs to the block.
Each PE in the tensor block can implement 1$\times$ \texttt{int16}, \texttt{fp16}, or 16-bit Brain floating-point (\texttt{bfloat16})~\cite{kalamkar2019study} MAC, and it can also be fractured to implement 4$\times$ \texttt{int8} MACs.

This tensor block has a 4.4$\times$ higher area footprint than an Intel Agilex-like DSP block, with 2.4$\times$ and 4$\times$ more input and output pins interfacing with the programmable routing, respectively.
To accommodate their higher area and increased signal demand, these blocks occupy multiple locations in the \emph{grid} defined by the FPGA routing channels. 
Hence they can connect to multiple routing channels; a single block spans 8 rows of the FPGA grid. 
A tensor block column is also physically 3.5$\times$ wider than an LB column. 
On a set of 9 DL benchmarks, the addition of these in-fabric tensor blocks increased the maximum operating frequency by 65\% and decreased the routed wirelength by 55\% on average. 
A large number of MAC operations and the interconnect between them can be mapped to the PEs of a single tensor block, leading to these speed and wirelength gains over distributed LBs and DSPs connected together using the programmable routing.
For non-DL benchmarks the tensor blocks not only remain idle but also, due to their coarse granularity, force other circuit components to be placed physically further away from each other with longer connections between them.
This results in a 0.5-2.5\% degradation in frequency and a 2-8\% increase in routed wirelength as the portion of die area dedicated for tensor blocks is varied from 5\% to 30\%. 

\begin{figure}[!t]
\centering
\includegraphics[width=\linewidth]{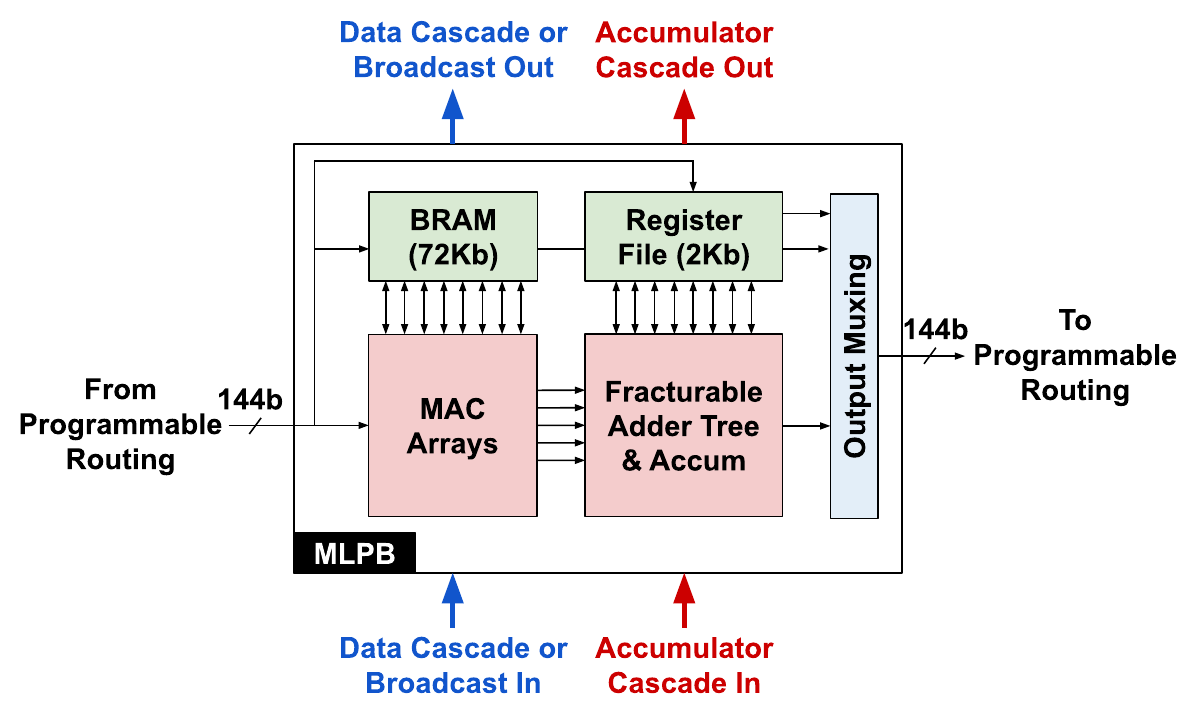}
\caption{The Achronix Speedster7t machine learning processor block (MLPB) internal architecture. Tightly coupling BRAMs and register files with MAC arrays restricts the high-bandwidth data transfers internally in the MLPB and limits the number of external interfaces to the programmable routing.}
\label{fig:mlpb}
\end{figure}

As the market for FPGA-based DL acceleration continues to grow rapidly, several FPGA vendors have also started to offer DL-optimized FPGA families that integrate different forms of in-fabric tensor blocks.
These devices sacrifice backward compatibility by entirely replacing the wireless-communication-targeted conventional DSP blocks  with new tensor blocks optimized specifically for the compute patterns and numerical precisions of DL workloads.
The Achronix Speedster7t FPGA~\cite{speedster7tmlp} integrates machine learning processor blocks (MLPBs\footnote{Although Achronix abbreviates their machine learning processor blocks as MLPs, we use MLPBs to avoid confusion with MLPs for multi-layer perceptron models.}), that tightly couple BRAMs and MAC units with dedicated high-bandwidth routing between them, as shown in Fig.~\ref{fig:mlpb}.
This tight coupling reduces the number of expensive interfaces to the programmable routing needed to feed the compute units inside the block.
New weight and/or activation data can be written to a double-buffered internal BRAM using relatively narrower external interfaces, while another set of weights and/or activations is reused for many compute operations with wide internal dedicated connections between the BRAM and compute units.

Another key benefit of this tight coupling of BRAMs and MAC units is that it enables these hard MLPBs to operate on a higher frequency clock domain (up to 750MHz) than the rest of the design implemented in soft logic, without the need to use the (slower and less efficient) fine-grained programmable routing for transporting data between memory and compute as in conventional FPGA fabrics. 
These MLPBs also natively support a wide variety of numerical precisions suitable for DL training and inference such as \texttt{int4/8/16}, \texttt{bfp12/16}, \texttt{bfloat16}, and \texttt{fp16/24}.
The largest Speedster7t devices include 2,560 MLPBs that can provide up to 61.4 and 122.8 TOPS of \texttt{int8/bfp16} and \texttt{int4/bfp12} performance, respectively.
Cairncross et al.~\cite{cairncrossai2023ai} demonstrated the use of the Speedster7t MLPBs to implement a 4-core FPGA DL overlay for low-latency inference use cases. 
The overlay can clock the MLPBs at 560MHz, achieving a peak \texttt{int8} performance of 36.4 TOPS with 80-100\% utilization of the compute units across a variety of GEMV, MLP, and RNN workloads at a batch size of 4.

\begin{figure}[!t]
\centering
\includegraphics[width=0.8\linewidth]{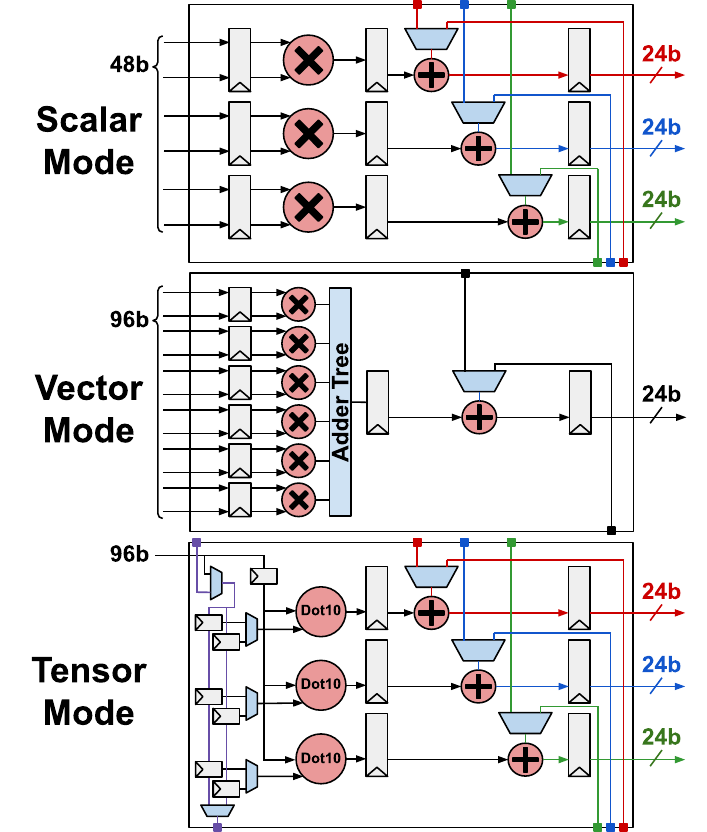}
\caption{Different \texttt{int8} modes of operation of the Intel Stratix 10 NX AI tensor block: scalar mode with 3 independent MACs (top), vector mode with one dot-6 operation without input restrictions (middle), and tensor mode with three dot-10 operations using input broadcast and input reuse register chains (bottom).}
\label{fig:aitb}
\end{figure}

The artificial intelligence tensor blocks (AITBs) in the Intel Stratix 10 NX device~\cite{langhammer2022stratix} are another example of commercial in-fabric tensor compute.
Although the end goal is the same (to integrate in-fabric tensor compute for DL), Intel adopted a different design approach from the academic tensor blocks and the Achronix MLPBs. 
The AITBs were designed as a drop-in replacement for conventional Stratix 10 DSP blocks in terms of silicon area footprint and interfaces to the programmable routing (i.e. only the block internals are different).
A single AITB has enough silicon footprint to implement up to 30 \texttt{int8} or 60 \texttt{int4} multipliers.
However, this would require 480 input and 480 output interfaces to the programmable routing, which is much higher (and would be much larger) than the 96 inputs and 72 outputs in the conventional Stratix 10 DSP block. 
Most DL workloads are dominated by operations where the results of many multiplies are accumulated and there is re-use of input data. Intel exploits this by designing three different AITB modes (shown in Fig.~\ref{fig:aitb}) that enable different levels of arithmetic density while staying within the 96 input / 72 output limit; more dense modes support increasingly constrained (but useful in DL) compute patterns.

In \emph{scalar mode}, the AITB performs completely independent multiplies. This mode is easy to use, but compute density is limited by the number of outputs to the general programmable routing; the AITB can perform only three independent \texttt{int8} MAC operations with a 24-bit accumulator each (i.e.~a total of 72 outputs).
The \emph{vector mode} internally sums its multiplies to produce one output, making it well suited for dot products. 
In this case, the AITB is limited by the number of inputs and can perform six \texttt{int8} multiplies in a dot-6 operation (i.e.~2~operands $\times$ 6~elements $\times$ 8~bits = 96 external inputs).
Finally, the \emph{tensor mode} provides the highest arithmetic density, but with more restrictions on the inputs and outputs of the AITB. To limit outputs to 72, it performs three \texttt{int8} dot-10 operations, each of which has an accumulator and dedicated interconnects to reduce results across AITBs in the same column. 
To limit inputs to 96, an input vector is broadcast to all 3 dot product units while three other input vectors (one per dot product unit) are fed locally by ping pong input reuse register chains. 
While inputs are reused for some time in many DL computations (e.g. by computing multiple output feature maps in a CNN from the same input maps), they must eventually be reloaded to proceed to the next set of inputs. 
This means that the AITB needs to either stall computations and load the reuse registers in parallel, or use the first block in a group of cascaded AITBs to sequentially load inputs to one of the reuse register chains using the dedicated AITB-to-AITB interconnect while the other chain is used for computation.
Additional lightweight circuitry is also added to the AITB to reuse the \texttt{int8} and \texttt{int4} multipliers for natively supporting \texttt{bfp16} and \texttt{bfp12} precisions, respectively.

\begin{figure}[!t]
\centering
\includegraphics[width=\linewidth]{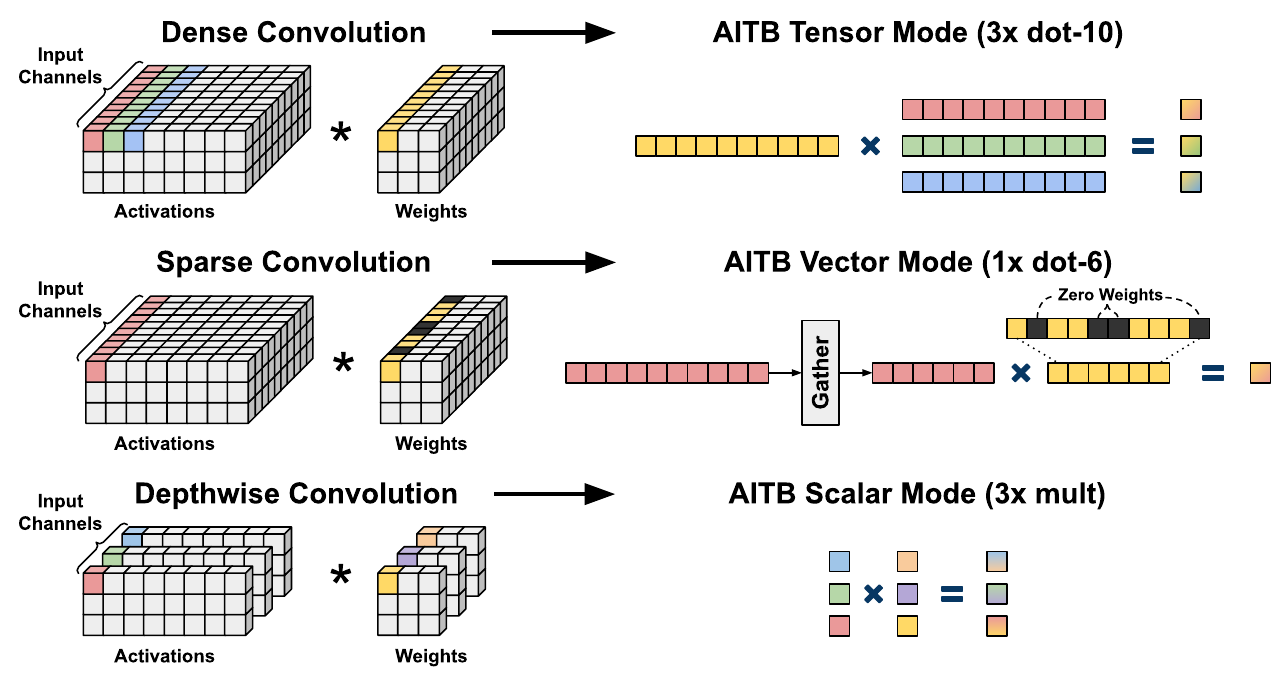}
\caption{Mapping of different convolution operations in HPIPE to different modes of the Intel Stratix 10 NX AITB modes of operation.}
\label{fig:hpipe_aitb}
\end{figure}

As discussed in Sec.~\ref{sec:tb_effect}, both the HPIPE~\cite{stan2022hpipe} and  NPU~\cite{boutros2020beyond} accelerators have been re-architected to make best use of the Intel Stratix 10 NX AITBs.
In HPIPE, all 3 modes of operation of the AITBs are used for different CNN operations as illustrated in Fig.~\ref{fig:hpipe_aitb}.
To exploit unstructured sparsity, HPIPE builds a multiplexing network in the soft logic to gather the activations matching non-zero weights. This maps well to the AITB vector mode due to its input flexibility (dot-6 operations with arbitrary inputs each cycle), and enables a 1.9$\times$ overall inference speedup compared to conventional DSP blocks. 
When running dense regular and pointwise convolutions, HPIPE can exploit the high arithmetic density of the tensor mode by pre-loading activations to the reuse register chains and broadcasting weights to all 3 dot units in the AITB.
However, the scalar mode remains necessary for implementing the depthwise convolutions as they do not have reduction or data reuse across the input channel dimension. 
Combining tensor and scalar modes speeds up HPIPE dense CNN inference by 5$\times$ compared to using conventional DSP blocks.
The NPU can also exploit the tensor mode of the AITBs, but in its case it is necessary to increase the batch size from 1 to 3. Activations from a batch of 3 different inputs are pre-loaded to the reuse register chains while weights are broadcast to the 3 dot product units.
This results in 3.5$\times$ higher throughput than the baseline NPU using DSP blocks. 

These performance gains all come with no increase in the FPGA die size, since the AITBs have the exact same area footprint and programmable routing interfaces as the DSP blocks they replace. 
While the gains are significant, they fall short of the 15$\times$ increase in peak \texttt{int8} TOPS compared to DSP blocks.
The peak performance can only be achieved if all operations match the compute pattern of the AITB tensor mode, all vector operands are a multiple of 10 elements to exactly fit the dot product units, and there is no overhead for loading data to the input reuse chains. One or more of these 3 requirements for ideal efficiency are not met in most application designs.
In addition, efficiently using the AITBs requires considerable changes to a design that originally targeted conventional DSP blocks; one cannot simply re-compile an RTL or HLS design to target these new AITBs.
The design computations must first be restructured to match one of the compute patterns supported by the different AITB modes, and then AITBs are instantiated as \emph{black-box} IPs in RTL to implement these computations. The resulting optimized designs are less portable between different FPGA families. 

Moving forward, the DSP blocks in the upcoming Intel Agilex 5 FPGA family will support modes from conventional DSP blocks (6$\times$\texttt{int9}, 2$\times$\texttt{int18}, 1$\times$\texttt{int27}), as well as a variation of the AITB tensor mode with only two (instead of three) \texttt{int8} dot-10 operations per block~\cite{agilex5}. Both the NPU and HPIPE require not only many low precision MAC operations but a few higher precision ones, so they will benefit from the ability to do both efficiently in one block. 
These hybrid blocks are targeted at edge DL applications in which the FPGA implements a full system where inference is a component along side other signal processing functionalities. 
\section{Beyond the FPGA Fabric}
\label{sec:out_of_fabric}

\begin{figure*}[!t]
\centering
\includegraphics[width=\linewidth]{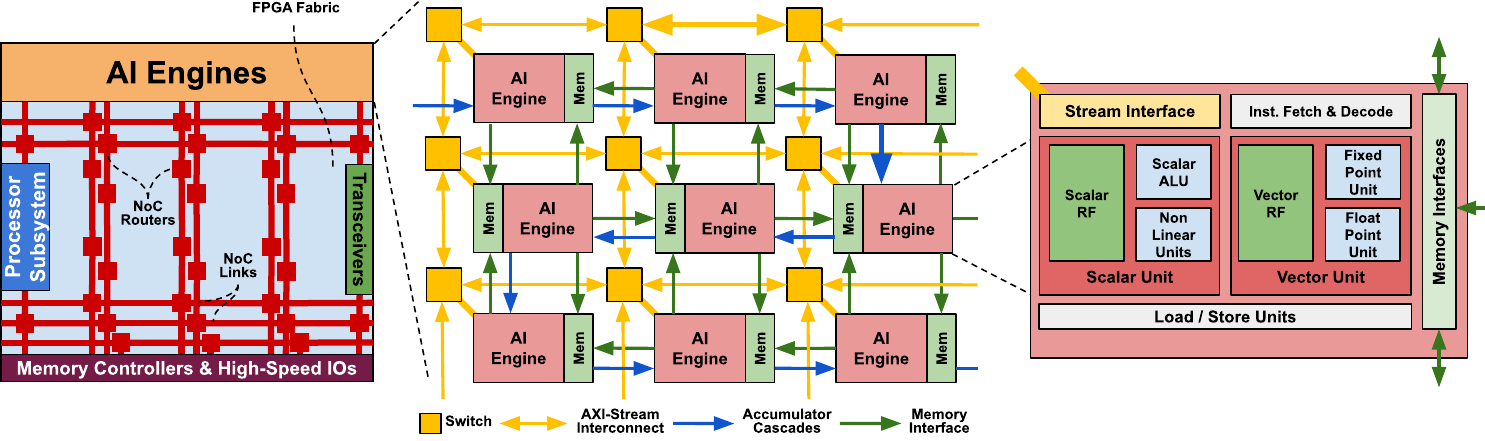}
\caption{The AMD Versal architecture combining an FPGA fabric, general-purpose processors, AI engines, and a packet-switched NoC with a modified mesh topology. The AI engines are arranged in a 2D grid with dedicated interconnect cascading their accumulators and an AXI packet/circuit-switched bus-based interconnect. Each AI engine is a VLIW vector processor with a peak throughput of 256 \texttt{int8} GOPS.}
\label{fig:versal}
\end{figure*}
Sections~\ref{sec:existing_blocks} and~\ref{sec:new_blocks} described DL-targeted enhancements to existing FPGA fabric components as well as the embedding of new tensor compute blocks in the fabric.
Beyond these improvements to the \emph{fine-grained} programmable fabric, several other architecture enhancements have been proposed to significantly increase peak performance by integrating \textit{coarse-grained} accelerator cores either on the same monolithic die or in package using advanced chip integration technologies.

\subsection{Reconfigurable Acceleration Devices}
Recently, a new class of \textit{reconfigurable acceleration devices} (RADs)~\cite{boutros2022rad} has emerged that combine the reconfigurability of conventional FPGA fabrics with the efficiency of coarse-grained application-specific accelerators and the flexibility of general-purpose processor cores.
These components are all connected via high-performance packet-switched networks-on-chip (NoCs) for system-wide communication.

One example of such a RAD is the AMD Versal architecture which tightly integrates a 7nm FPGA programmable fabric, general-purpose Arm Cortex cores, and a 2D array of specialized vector processors termed adaptive intelligent engines (AIEs) on the same monolithic die~\cite{gaide2019xilinx}.
These different system components as well as the modules on the FPGA fabric communicate using a hard packet-switched NoC. The NoC is also the only way to access external memories (e.g.~DDR or HBM).
The Versal NoC~\cite{swarbrick2019network} has a modified mesh topology were several columns are grouped together and rows are \textit{squished} to the top and bottom of the device as illustrated in Fig.~\ref{fig:versal}. 
This topology matches the columnar nature of the FPGA fabric, simplifies the layout of the chip, and also provides higher bandwidth for horizontal communication at the top and bottom of the device where the high-speed IOs, memory controllers, and the AIE array are located.
The presence of a hard NoC significantly boosts FPGA designer productivity and facilitates timing closure; 
it is no longer necessary to go through many design iterations to optimize the system-level interconnect built using the (relatively less efficient) programmable routing resources~\cite{abdelfattah2013case}. 
Different modules implemented on the programmable fabric and communicating via latency-insensitive interfaces can be independently and locally optimized to close timing as standalone components. 
Then the compiled modules can be connected to one of the pervasive NoC \emph{fabric ports} to communicate with other fabric modules, coarse-grained accelerators (e.g. AIEs), and external interfaces.
This can be extremely useful, especially for large and complex FPGA systems with many compute modules and high external memory bandwidth requirements such as DL acceleration designs.

In addition, the AIEs significantly enhance the compute capabilities of the Versal architecture by combining an array of vector processors with FPGA-like spatial interconnect and distributed state in a hybrid computational paradigm.
Each AIE tile contains a 1~GHz very-long-instruction-word (VLIW) vector processor that can execute 7 operations simultaneously (2 vector loads, 1 vector store, 1 vector operation, and 2 scalar operations).
The fixed-point vector unit in a single AIE is capable of performing 128 \texttt{int8} MAC operations per cycle for a peak throughput of 256 GOPS.
The vector processor is tightly coupled with 32KB of local SRAM memory and a direct memory access engine for non-neighbor data communication.
As illustrated in Fig.~\ref{fig:versal}, the AIE tiles are arranged in a 2D grid with an AXI-Stream interconnect network that can implement both circuit-switched and packet-switched communication between remote tiles.
In addition, there is a dedicated interconnect that cascades accumulators between neighbouring AIEs in a serpentine pattern; this interconnect is conceptually similar to the accumulation cascade chains between DSP blocks in a conventional programmable fabric. 
Each AIE also has the ability to directly read from and write to the local memory of 3 adjacent neighbours (north, south, and east or west depending on the physical layout of the SRAM memories as shown in Fig.~\ref{fig:versal}).
The biggest Versal device has an array of 400 AIEs that can provide more than 100 TOPS of \texttt{int8} compute, in addition to the compute units that can be implemented in the conventional FPGA fabric.

Several works have demonstrated the use of the Versal AIEs for accelerating different DL workloads, such as CNNs~\cite{jia2022xvdpu, zhang2022u3d}, transformer networks~\cite{zhuang2023charm}, and graph neural networks~\cite{zhang2022hgcn, chen2023exploiting}.
The use of such spatial coarse-grained cores running at a much higher frequency than the FPGA's programmable fabric can significantly improve DL inference efficiency.
However, efficiently mapping an application to a large number of software-programmable vector processors can be a challenging task.
Since the AIEs introduce a new reconfigurable acceleration paradigm, the CAD tools to support them are not yet mature; improving the quality of results the CAD flow can achieve with less designer intervention is an active area of research. 
Several works have introduced frameworks that can optimize the mapping of matrix-matrix multiplication kernels of different sizes and compositions to an AIE array such as CHARM~\cite{zhuang2023charm} and MaxEVA~\cite{taka2023maxeva}.
The Versal AIEs can also be used to accelerate other non-DL workloads that can benefit from their vector processing capabilities and spatial nature such as stencil-based scientific simulations~\cite{brown2023exploring, singh2023sparta}.

\begin{figure}[!t]
\centering
\includegraphics[width=\linewidth]{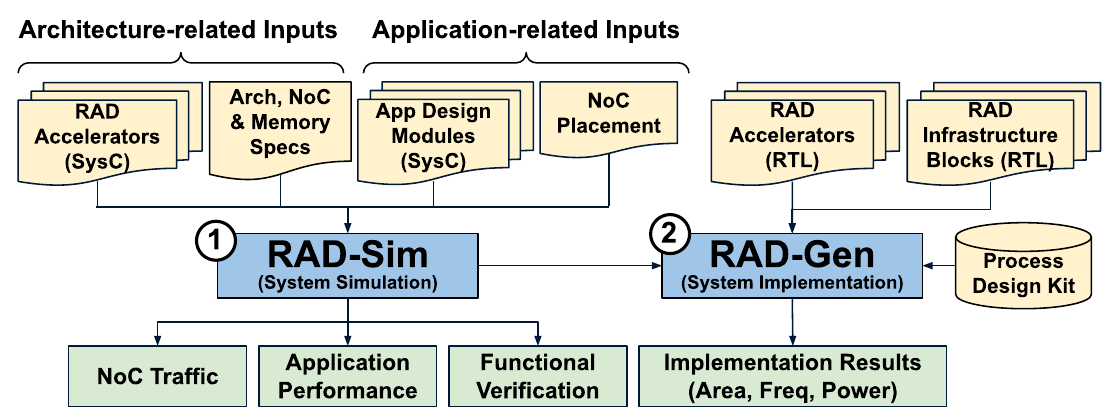}
\caption{Architecture exploration and evaluation flow for novel RADs.}
\label{fig:rad_flow}
\end{figure}

The AMD Versal architecture is one specific RAD instance from the huge design space arising from the combination of fine-grained programmable fabrics, coarse-grained accelerator cores, and NoC(s) for system-level interconnect. 
This vast design space remains little explored due to the dearth of tools that can model different RAD architectures and enable evaluation of the complex interactions between their various components.
The work presented in~\cite{boutros2022rad, boutros2022architecture, boutros2023whole} aims to fill this gap by introducing a complete architecture exploration and evaluation flow for RADs, as shown in Fig.~\ref{fig:rad_flow}.
The first component of this flow is RAD-Sim, a cycle accurate architecture simulator for RADs.
It provides the infrastructure for simulating RADs with different architecture parameters, NoC specifications, and hardened accelerator cores.
A user can define application modules to be implemented on the RAD FPGA fabric and/or coarse-grained accelerator cores in SystemC, connect them to the RAD NoC, and simulate the entire system. 
RAD-Sim can then report the end-to-end application performance and NoC traffic congestion, as well as verify the application functionality on a candidate RAD architecture. 
This can be used to rapidly explore the design of both RAD architectures and applications~\cite{boutros2022architecture}.
These RAD devices introduce a new placement problem: to which physical NoC router should each accelerator or programmable logic module connect? 
Architects can either experiment with different placements manually, or RAD-Sim can interact with the VTR placement engine to automatically determine an optimized NoC placement~\cite{srinivasan2023placement}.

After the design space is narrowed down to a few RAD candidates, the second component of this flow, RAD-Gen, can be used to evaluate their implementation feasibility and cost.
For example, an accelerator core might significantly improve performance in terms of cycle count when modeled in RAD-Sim, but might not fit within its silicon area budget or might run at a slower frequency than assumed.
RAD-Gen takes as inputs RTL descriptions of common RAD components (e.g. NoC routers) and/or candidate accelerator cores with a list of parameter values to be swept, as well as high-level ASIC implementation flow configurations and the design kit for the target process.
Then, it automatically runs the ASIC synthesis, place and route, and timing analysis tools to evaluate the area and performance of different variations of these RAD components.
By using both RAD-Sim and RAD-Gen, an architect can evaluate the performance-cost tradeoff for different RAD candidates as demonstrated for DL recommendation model acceleration in~\cite{boutros2023whole}.

\subsection{DL Chiplets}

\begin{figure}[!t]
\centering
\includegraphics[width=\linewidth]{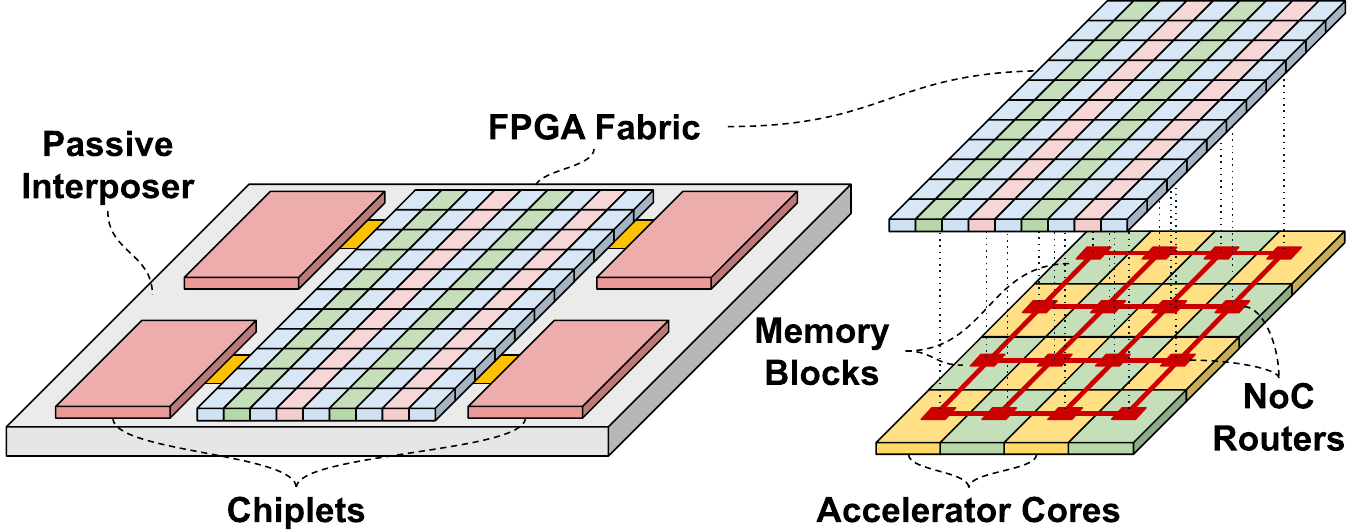}
\caption{Passive interposers for integrating an FPGA fabric with accelerator chiplets (left) and 3D stacked RADs (right).}
\label{fig:chiplets_and_3d}
\end{figure}

Most modern FPGAs from AMD and Intel have been using interposer technology to integrate either multiple FPGA dice or an FPGA die and one or multiple I/O transceiver \emph{chiplets} in the same package with high interconnect density between them, as depicted in Fig.~\ref{fig:interposer}.
Interposers can also be used to build FPGA devices targeting a certain application domain by integrating a specialized ASIC chiplet in the same package, as depicted in the left side of Fig.~\ref{fig:chiplets_and_3d}.
Work from Intel Labs proposed the integration of different accelerator chiplets in the Stratix 10 FPGA package to enhance DL inference efficiency for different workloads~\cite{nurvitadhi2018package, nurvitadhi2019compete}.
In this case, the chiplet implements tensor operations that are common across a wide variety of DL workloads, freeing up FPGA fabric resources to implement either model layers that can change over time (e.g. residual connections or activation functions) or other system components such as pre/post-processing stages.
Nurvitadhi et al.~\cite{nurvitadhi2019compete} integrated TensorRAM, a chiplet optimized for memory-bound DL models, with a small Stratix 10 FPGA in the same package to achieve 16$\times$ lower latency and 34$\times$ higher energy efficiency compared to the largest same-generation Nvidia DL-optimized GPU.

More recent advances in chip integration technology have enabled the stacking of multiple active dice on top of each other~\cite{ingerly2019foveros}.
For example, the announced Instinct MI300 datacenter GPU accelerator family from AMD uses active die stacking technology to integrate 13 chiplets including CPU and GPU cores on top of dice that handle IO and memory traffic~\cite{su2023ceskeynote}.
This also opens the door for a myriad of possibilities for 3D RADs that integrate an FPGA fabric on top of an ASIC base die that implements larger on-chip memories, application-specific accelerators for DL, and system-level NoCs, as shown in Fig.~\ref{fig:chiplets_and_3d}~\cite{boutros2023into}.
\section{Summary}
\label{sec:summary}

With DL becoming the cornerstone of a myriad of applications running on large-scale datacenter clusters as well as edge devices, there is a pressing need for efficient compute platforms that can keep up with the growing compute demands of DL models.
This has driven architectural innovations for general-purpose CPUs and GPUs and the creation of myriad ASIC DL accelerators.
FPGAs offer several unique features compared to these other compute platforms: 
(1) their fine-grained hardware programmability enables customizing numerical precisions and the on-chip memory hierarchy to exactly match the needs of the target DL models, (2) their spatial architecture can exploit massive parallelism and direct communication between compute units for inference applications with tight latency constraints, (3) their reconfigurability allows adding or changing hardware features as DL models evolve, and (4) their diverse IOs enable building end-to-end DL systems in which an inference component is interfaced with different sensors and actuators in edge applications or high-speed networking in datacenters.

In this article, we described different design styles of DL accelerators on FPGAs that achieve state-of-the-art performance while improving ease-of-use for application developers. 
Similarly to other compute platforms such as CPUs and GPUs, FPGA architecture is also evolving to better suit DL workloads.
We surveyed different proposals on how to enhance the FPGA underlying architecture to be even better at DL.
These enhancements include modifying conventional FPGA fabric blocks (logic blocks, DSPs, BRAMs), adding new in-fabric blocks for tensor compute, and integrating conventional FPGA fabrics with different coarse-grained accelerator cores and chiplets in future RADs.

The design space of RAD architectures is very large, as it comprises fabric optimizations, new coarse-grained accelerator blocks, and different methods to interconnect them using traditional programmable routing, NoCs, and 2.D or 3D integration. 
We expect exploration of these architectures to improve DL inference efficiency  will remain a dynamic research area for years to come.

\bibliographystyle{IEEEtran}
\bibliography{references}

\newpage

\section{Biography Section}

\begin{IEEEbiography}[{\includegraphics[width=1in,height=1.25in,clip,keepaspectratio]{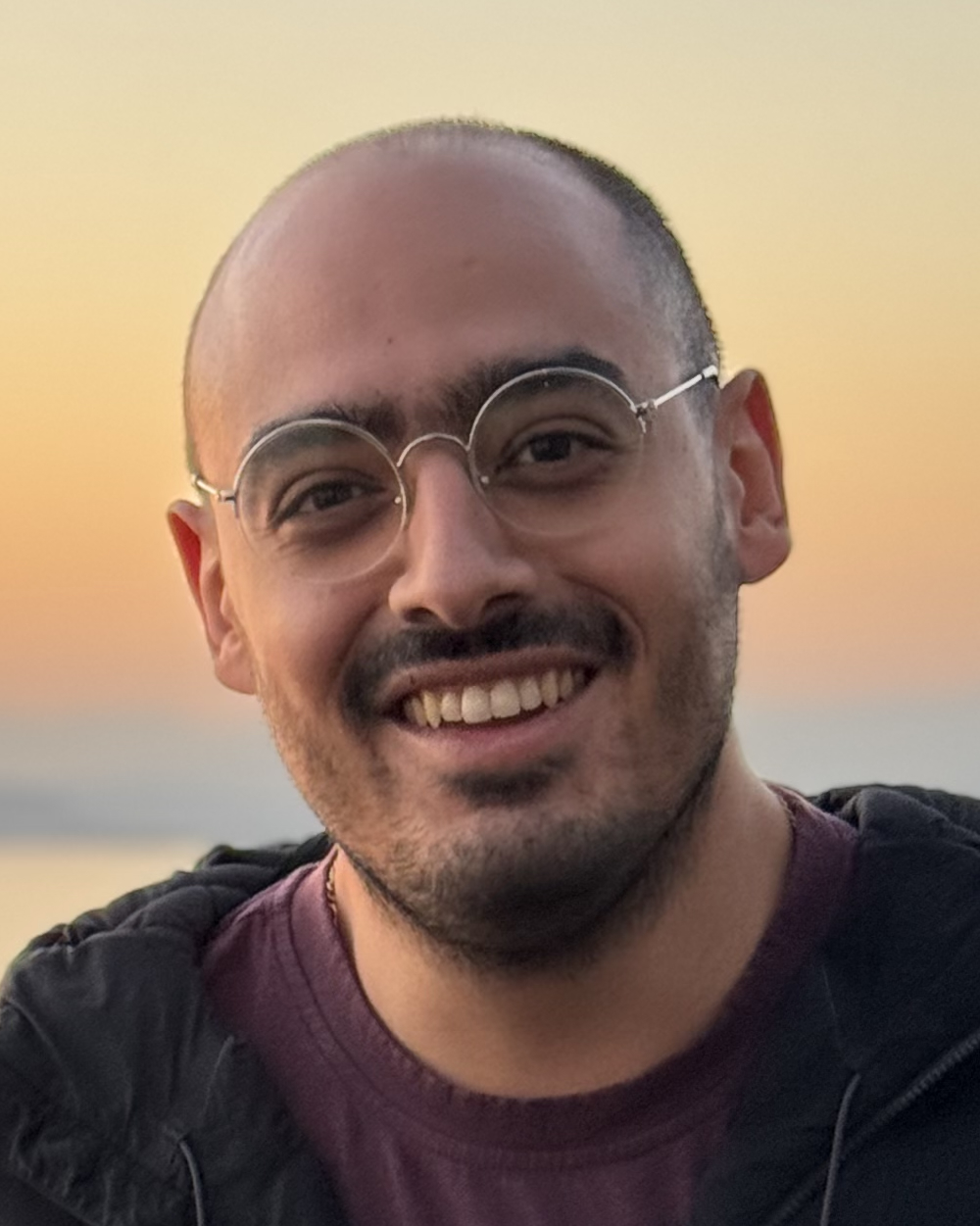}}]{Andrew Boutros}
received his B.Sc. degree in electronics engineering from the German University in Cairo in 2016, and his M.A.Sc. degree in electrical and computer engineering from the University of Toronto in 2018. He was a research scientist at Intel’s Accelerator Architecture Lab in Oregon before returning to the University of Toronto, where he is currently pursuing his Ph.D. degree under the supervision of Prof. Vaughn Betz. His research interests include FPGA architecture and CAD, deep learning acceleration, and next-generation reconfigurable acceleration devices. He is an affiliate of the Intel/VMware Crossroads 3D-FPGA Academic Research Center and the Center for Spatial Computational Learning. He has more than 30 publications in top conferences and journals in the field of FPGAs, including 4 best paper awards. 
\end{IEEEbiography}

\begin{IEEEbiography}[{\includegraphics[width=1in,height=1.25in,clip,keepaspectratio]{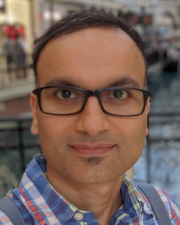}}]{Aman Arora} received his B.Tech. degree in electronics and communications engineering from the National Institute of Technology Kurukshetra in 2007, and his M.S. and Ph.D. in electrical and computer engineering from the University of Texas at Austin in 2012 and 2023, respectively.
He is currently an assistant professor at Arizona State University, where he leads a research lab that focuses on next-generation FPGA architectures and hardware for machine learning. During his graduate school, he focused on optimizing FPGA architecture for deep learning workloads. He has received a best paper at FCCM 2022 and has over 10 years of experience in the semiconductor industry in design, verification, testing and architecture roles. 
\end{IEEEbiography}

\begin{IEEEbiography}[{\includegraphics[width=1in,height=1.25in,clip,keepaspectratio]{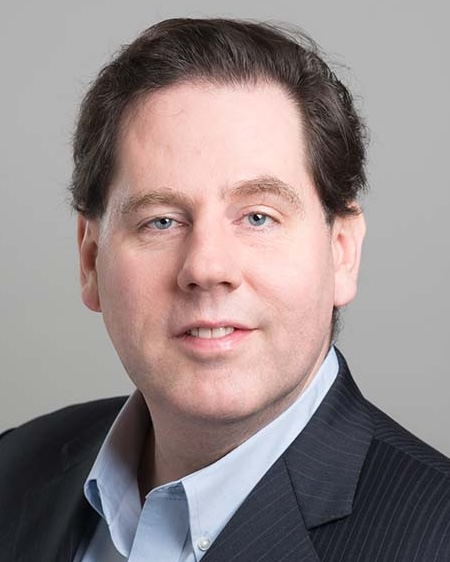}}]{Vaughn Betz} received his B.Sc. degree in electrical engineering from the University of Manitoba in 1991, his M.S. degree in electrical and computer engineering from the University of Illinois at Urbana–Champaign in 1993, and his Ph.D. degree in electrical and computer engineering from the University of Toronto in 1998. He is the original developer of the widely used VPR FPGA CAD flow. He co-founded Right Track CAD to commercialize VPR, and joined Altera upon its acquisition of Right Track CAD. He spent 11 years at Altera, ultimately as Senior Director of software engineering, and is one of the architects of the Quartus CAD system and the first five generations of the Stratix and Cyclone FPGA families. He is currently a professor and the NSERC/Intel Industrial Research Chair in Programmable Silicon at the University of Toronto. He holds 102 US patents and has published over 100 technical articles in the FPGA area, sixteen of which have won best or most significant paper awards. He is a Fellow of the IEEE, the ACM, the National Academy of Inventors, and the Engineering Institute of Canada. He is also a Faculty Affiliate of the Vector Institute for Artificial Intelligence.
\end{IEEEbiography}

\vfill

\end{document}